\documentclass[%
repsrint,
twocolumn,
superscriptaddress,
showpacs,
amsmath,amssymb,amsfonts, 
aps,
pre,
floatfix,notitlepage,latexsym%,longbibliography
]{revtex4-1}

\usepackage[dvipdfmx]{graphicx}% Include figure files
\usepackage{dcolumn}% Align table columns on decimal point
\usepackage{bm}% bold mathematica
\usepackage{braket}
\usepackage[dvipdfmx]{attachfile2}
\usepackage{subfigure}
\usepackage{comment}

\newcommand{\Exp}[1]{\mathrm{e}^{#1}}
\newcommand{\mL}{\mathcal{L}}
\newcommand{\mV}{\mathcal{V}}
\newcommand{\tr}{\text{tr}}

\newcommand{\oo}{O(\omega^{-1})}
\newcommand{\ot}{O(\omega^{-2})}

\newcommand{\hzero}{\hat{H}_0}

\newcommand{\htot}{\hat{H}}
\newcommand{\hrho}{\hat{\rho}}
\newcommand{\lind}{\hat{L}}

\newcommand{\hnv}{\hat{H}_\text{NV}}
\newcommand{\hH}{\hat{H}}

\newcommand{\Heff}{\hat{H}_{\mathrm{eff}}}
\newcommand{\dHeff}{\varDelta\hat{H}_{\mathrm{eff}}}

\newcommand{\Leff}{\mathcal{L}_\text{eff}}
\newcommand{\mG}{\mathcal{G}}
\newcommand{\mD}{\mathcal{D}}

\newcommand{\hO}{\hat{A}}

\newcommand{\hSx}{\hat{S}_x}
\newcommand{\hSy}{\hat{S}_y}
\newcommand{\hSz}{\hat{S}_z}	
\newcommand{\hUz}{\hat{U}^z}
\newcommand{\hUzd}{\hat{U}^{z\dag}}
\newcommand{\hUzt}{\hat{U}^z_{\theta}}
\newcommand{\hUztd}{\hat{U}^{z\dag}_{\theta}}

\newcommand{\hUzp}{\hat{U}^{z}_\pi}
\newcommand{\hUzpd}{\hat{U}^{z\dag}_\pi}

\newcommand{\rhoss}{\hat{\rho}_\text{ness}}

\newcommand{\rhoinf}{\hat{\rho}'_\infty}
\newcommand{\rhoinfel}[1]{\rho_{\infty}^{\prime\,{#1}}}
\newcommand{\rhoinfd}{\rho_{\infty}^{\prime\,(\text{d})}}
\newcommand{\rhoinfoff}{\rho_{\infty}^{\prime\,(\text{od})}}

\newcommand{\rhoUzp}{\hat{\rho}^{\hUzp}}
\newcommand{\rhossUzp}{\hat{\rho}_{\text{ness}}^{\hUzp}}
\newcommand{\rhoV}{\hat{\rho}^{\hat{V}}}

\newcommand{\cfss}{\hat{\rho}_\text{CFSS}}
\newcommand{\fgs}{\hat{\rho}_\text{FG}}

\newcommand{\rhocan}{\hrho_\text{can}}
\newcommand{\sigmm}{\hat{\sigma}_{\text{MM}}}
\newcommand{\sigfe}{\hat{\sigma}_{\text{FE}}}
\newcommand{\canw}{p_{\text{can}}}
\newcommand{\fgw}{p_{\text{FG}}}

\newcommand{\rhofs}{\hat{\rho}^{\text{FS}}}

\newcommand{\tket}[1]{\widetilde{\ket{#1}}}

\newcommand{\ad}{\text{ad}}

\begin{abstract}
Laser technology has developed and accelerated photo-induced nonequilibrium physics from both scientific and engineering viewpoints. The Floquet engineering, i.e., controlling material properties and functionalities by time-periodic drives, is a forefront of quantum physics of light-matter interaction, but limited to ideal dissipationless systems. For the Floquet engineering extended to a variety of materials, it is vital to understand the quantum states emerging in a balance of the periodic drive and energy dissipation. Here we derive the general description for nonequilibrium steady states (NESS) in periodically driven dissipative systems by focusing on the systems under high-frequency drive and time-independent Lindblad-type dissipation with the detailed balance condition. Our formula correctly describes the time-average, fluctuation, and symmetry property of the NESS, and can be computed efficiently in numerical calculations. Our approach will play fundamental roles in Floquet engineering in a broad class of dissipative quantum systems such as atoms and molecules, mesoscopic systems, and condensed matters.
\end{abstract}

\begin{document}

\title{General description for nonequilibrium steady states\\in periodically driven dissipative quantum systems}
\author{Tatsuhiko N. Ikeda}
\email{tikeda@issp.u-tokyo.ac.jp}
\affiliation{Institute for Solid State Physics, University of Tokyo, Kashiwa 277-8581, Japan}
%\author{Sho Higashikawa}
%\affiliation{Department of Physics, University of Tokyo, Hongo 113-8656, Japan}
\author{Masahiro Sato}
\email{masahiro.sato.phys@vc.ibaraki.ac.jp}
\affiliation{Department of Physics, Ibaraki University, 
Mito, Ibaraki 310-8512, Japan}
\date{\today}

\maketitle

\section{Introduction}
State-of-the-art laser technology
has opened new research fields in physics: the Floquet science and engineering~\cite{Holthaus2015, BukovM15, Oka2019}.
The main focus of these fields is the nonequilibrium states driven periodically by external fields, e.g., intense laser fields.
Physical properties of the nonequilibrium states
are mainly understood by the so-called effective Hamiltonian,
which reflects the periodic driving, according to the Floquet theorem~\cite{ShirleyJ65} and the ensuing theoretical developments~\cite{EckardtA15, MikamiT16,LazaridesA14,KuwaharaT16}.
Conversely, designing a suitable driving protocol,
one can \textit{engineer} the effective Hamiltonian,
which enables us to have desirable properties and functionalities of physical systems.
Indeed, various exotic states and useful manipulation of matter have been theoretically proposed and 
some of them have been experimentally realized: 
Floquet topological states \cite{OkaT09} in solids \cite{WangY13}, 
ultracold atomic gases \cite{JotzuG14}, and in photonic wave guides \cite{RechtsmanM13}, 
Floquet time crystals \cite{ElseD16} in nitrogen-vacancy centers \cite{ChoiS17} 
and trapped ions \cite{ZhangJ17}, and control of quantum magnets~\cite{Sato2016} and their interactions~\cite{Mentink2015}.

However, these Floquet-theoretical predictions based on the effective Hamiltonian are quantitatively only in ultraclean materials or well-designed artificial systems, where dissipation is negligible. For the Floquet science and engineering in real generic materials,
it is indispensable to understand the nonequilibrium steady state (NESS),
which emerges in a balance of the energy injection by the periodic driving and the energy dissipation~\cite{KohnW01, HoneD09, KohlerS97, BreuerH00}.
For individual systems, by considering specific sources of dissipation i.e. system-bath couplings, one can calculate physical quantities in the NESS and predict interesting phenomena such as Floquet topological insulators~\cite{DehghaniH14,SeetharamK15}, periodic thermodynamics~\cite{Schmidt2019a}, dynamical localization~\cite{BlumelR91}, and generalized Bose-Einstein condensation~\cite{VorbergD13}. In this research direction, the Floquet-Green-function approach has developed and enables us to calculate various physical effects dependent on the type of the system-bath coupling~\cite{Kohler2004,Stefanucci2008}. Another research direction, which we address here, is to seek for a universal characterization for the NESS. We could imagine that there exists a simple and general expression for the NESS when the dissipation is weak and featureless. An
attempt is to conjecture that the NESS is generally described by the Floquet-Gibbs state (FGS),
i.e., the Gibbs state with the effective Hamiltonian,
but the conditions for the FGS being realized have shown quite restrictive~\cite{ShiraiT15, ShiraiT16, LiuD15}.
Hence, despite its importance,
the general formula for the NESS has been still an elusive problem.

In this paper, in exchange for restricting ourselves to the high-frequency drivings,
we deal with generic systems and driving protocols,
obtaining simple and general formulas for the NESS [Eqs.~\eqref{eq: MR1}--\eqref{eq: FE} below].
We obtain these formulas by applying the high-frequency expansion technique,
which has been recently developed~\cite{GoldmanN14a, RahavS03, EckardtA15, MikamiT16,DaiC16},
to the Lindblad equation with periodic Hamiltonians.
As exemplified in an effective model for the NV center in diamonds~\cite{Rondin2014},
our formulas correctly describe both the time average and fluctuation
of the NESS at the leading order of $\omega^{-1}$ ($\omega$ denotes the driving frequency).
These formulas also capture
nontrivial behaviors of physical quantities
due to the dynamical-symmetry breaking
that cannot be described by the effective Hamiltonian or the FGS,
and will thereby play critical roles in the Floquet science and engineering in dissipative quantum systems.

\section{Formulation of the problem}
We begin by considering a quantum system defined on an $N$-dimensional Hilbert space.
This system can be single-body or many-body as long as it satisfies the requirements that will be described below.
We let $H_0$ denote the time-independent Hamiltonian,
which describes our system in the absence of driving.
The eigenenergies and eigenstates of $H_0$ are denoted by $\{ E_i \}_{i=1}^N$ and $\{ \ket{E_i} \}_{i=1}^N$, respectively.
For simplicity, we assume that the eigenenergies are not degenerate
and $E_1< E_2< \cdots < E_N$
(the generalization
to degenerate $H_0$
is formulated in Supplemental Material).
The effect of the driving is represented by
a time-dependent part $H_{\text{ext}}(t)$ of the total Hamiltonian,
\begin{align}\label{eq: Htot}
    H (t) = H_0 + H_{\text{ext}}(t).
\end{align}
We assume that the driving term is periodic with period $T$: $H_{\text{ext}}(t+T)=H_{\text{ext}}(t)$ and hence $H(t+T)=H (t)$.
Without loss of generality, 
the decomposition~\eqref{eq: Htot} is defined so that
the time average of $H_{\text{ext}}(t)$ vanishes, $\int_0^T dt \ H_{\text{ext}} (t) = 0$. 
Thus the Fourier series of $H_{\text{ext}}(t)$ can be written as
\begin{align}
    H_{\text{ext}}(t) = \sum_{m\neq0} {H}_m e^{-i m \omega t}.
\end{align}

To study driven dissipative systems,
we consider 
the density operator $\rho(t)$
whose dynamics is described by the Lindblad equation \cite{Ho1986a,Prosen2011,Hartmann2017a,BreuerH02} (we set $\hbar=1$ throughout this paper):
\begin{align} \label{eq: Lindblad}
\frac{d \rho (t)}{dt} & = \mathcal{L}_t \rho(t)=- i \left[ H (t) , \rho (t) \right]+\mathcal{D}[\rho(t)],\notag\\
\mathcal{D}[\rho(t)]&\equiv
\sum_{i,j} 
\Gamma_{ij} \left(
L_{ij} \rho (t) L_{ij}^\dagger - 
\frac{1}{2} \left\{ L_{ij}^\dagger L_{ij}, \rho (t) \right\}
\right).
\end{align}
Here $L_{ij} := \ket{E_i}\bra{E_j}$ is
the time-independent Lindblad operator describing the transition
from the $j$-th to the $i$-th eigenstates of the undriven Hamiltonian $H_0$.
When $E_i< E_j$, $L_{ij}$ represents a decay (excitation) process for $i<j$ ($i>j$).
The real number $\Gamma_{ij}$ $(\ge0)$ denotes the rate for the corresponding process,
and we set $\Gamma_{ii}=0$ for each $i$.
The transition rates $\Gamma_{ij}$ must be small enough for the Floquet-Lindblad equation being valid (see Discussion below).
Note that Eq.~\eqref{eq: Lindblad} is trace-preserving $d\, \text{tr}[\rho (t)]/dt=0$,
and thus we use the normalization $\mathrm{tr}[\rho(t)]=1$.

We assume that the transition rates $\Gamma_{ij}$ satisfy the detailed balance condition,
\begin{align}\label{eq: DB}
\Gamma_{ij}e^{-\beta E_j} = \Gamma_{ji}e^{-\beta E_i} \qquad (\text{for}\ i\neq j),
\end{align}
where $\beta$ is the inverse temperature of the bath
coupled to the system (see Discussion below for generalization in the absence of this assumption).
We also assume that the matrix $\Gamma_{ij}$ is a nonnegative irreducible
matrix~\cite{Schmidt2019}.
These assumptions ensure that,
without driving,
the system goes, irrespective of the initial state, to 
the thermal equilibrium state,
or the canonical ensemble $\rho_\text{can}=e^{ - \beta H_0}/Z$ of $H_0$ with $Z=\text{tr}(e^{-\beta H_0})$.
We note that the Lindblad operators $L_{ij}$ may depend on 
the driving in general
if we consider more microscopic theories of dissipation~\cite{BreuerH00}.
However, we neglect this dependence in this work for simplicity.

\section{Derivation of main results by high-frequency expansion}
The key idea to obtain the nonequilibrium steady state is the high-frequency expansion for the Lindblad equation~\cite{DaiC16}.
Among several formulations,
we adopt the van Vleck perturbation theory~\cite{EckardtA15,MikamiT16},
which leads to the following propagation for $\rho(t)$
(see Supplementary Material for detail):
$\rho(t) = e^{\mathcal{G}(t)}e^{(t-t')\mathcal{L}_\text{eff} }e^{-\mathcal{G}(t')}\rho(t')$.
The time-independent part $\mathcal{L}_\text{eff}$ is represented by
the effective Hamiltonian 
\begin{align}
    \mathcal{L}_\text{eff}(\rho) = -i [{H}_{\mathrm{eff}},\rho] + \mathcal{D}(\rho)+O(\omega^{-2})
\end{align}
with ${H}_{\mathrm{eff}}=H_0 +\omega^{-1}\sum_{n>0}[{H}_{-n},{H}_n]/n+O(\omega^{-2})$.
The time-dependent part $e^{\mathcal{G}(t)}$
is the so-called micromotion operator
periodic in time $\mathcal{G}(t+T)=\mathcal{G}(t)$,
and given by $\mathcal{G}(t)(\rho)=\omega^{-1}\sum_{m\neq0}[{H}_m,\rho]e^{-im\omega t}/m+O(\omega^{-2})$.
Without loss of generality,
we suppose the initial time to be $t'=0$,
having
\begin{align}
\rho(t)=e^{\mathcal{G}(t)}e^{t\mathcal{L}_\text{eff} }e^{-\mathcal{G}(0)}\rho(0)
\end{align}
with $\rho(0)$ being our initial state.

To obtain the asymptotic behavior of $\rho(t)$,
we focus on the first part $\rho'(t)=e^{t\mathcal{L}_\text{eff} }e^{-\mathcal{G}(0)}\rho(0)$.
Remark that this is the solution of the time-independent
Lindblad equation $d\rho'(t)/dt=\mathcal{L}_\text{eff}\rho'(t)$
from the initial state $e^{-\mathcal{G}(0)}\rho(0)$.
Under our assumptions on $\mathcal{D}$, $\rho'(t)$ approaches,
irrespective of the initial state,
the unique state ${\rho}'_\infty$ characterized by
$\mathcal{L}_\text{eff}{\rho}'_\infty=0$.%~\footnote{Explain why this is unique}.
Thus we come to the first main result, obtaining the asymptotic behavior
\begin{align}\label{eq: MR1}
    \rho(t)\to{\rho}_\text{ness}(t) = e^{\mathcal{G}(t)}{\rho}'_\infty \qquad \text{as}\quad t\to\infty.
\end{align}
Since $\mathcal{G}(t)=\mathcal{G}(t+T)$,
this nonequilibrium steady state is also periodic in time.
Focusing on the leading-order contribution,
we have a simple explicit formula for ${\rho}_\text{ness}(t)$:
\begin{equation}\label{eq: MR2}
    {\rho}_\text{ness}(t)=\rho_\text{can}+{\sigma}_{\text{MM}}(t)+{\sigma}_{\text{FE}}+O(\omega^{-2}),
\end{equation}
in which both ${\sigma}_{\text{MM}}(t)$ and ${\sigma}_{\text{FE}}$ are $O(\omega^{-1})$ 
and we call ${\sigma}_{\text{MM}}(t)$ and ${\sigma}_{\text{FE}}$
the micromotion and Floquet engineering parts, respectively.
Equation~\eqref{eq: MR2} is our second main result,
which we prove in the Supplemental Material.
Its generalization in the absence of the detailed balance condition is outlined in Discussion below.
Note that $\text{tr}[{\rho}_\text{ness}(t)]=1$ is satisfied, at least, up to this order
since both ${\sigma}_{\text{MM}}(t)$ and ${\sigma}_{\text{FE}}$ are traceless as will be evident below.

The micromotion part ${\sigma}_{\text{MM}}(t)$ in Eq.~\eqref{eq: MR2}
is defined by
\begin{equation}\label{eq: MM}
    {\sigma}_{\text{MM}}(t) = \frac{1}{\omega}\sum_{m\neq0}\frac{e^{-im\omega t}}{m}[H_m,\rho_\text{can}].
\end{equation}
We have named it after the following two properties of ${\sigma}_{\text{MM}}(t)$.
First, this part is periodic in time ${\sigma}_{\text{MM}}(t+T)={\sigma}_{\text{MM}}(t)$
and contributes to oscillations of physical observables.
Second, ${\sigma}_{\text{MM}}(t)$ does not contribute to the time averages
of physical observables for one period of oscillations.
In fact, for an observable ${A}$,
we have $\int_{t_1}^{t_1+T}dt\,\text{tr}[{\sigma}_{\text{MM}}(t){A}]/T=0$.

The Floquet-engineering part ${\sigma}_{\text{FE}}$ in Eq.~\eqref{eq: MR2}
is independent of time and given by
\begin{equation}\label{eq: FE}
    \braket{E_k|{\sigma}_{\text{FE}}|E_l} =  
    \frac{\braket{E_k|\varDelta{H}_{\mathrm{eff}}|E_l}}{(E_k-E_l)-i\gamma_{kl}}
    (p_{\text{can}}^{(k)}-p_{\text{can}}^{(l)})
\end{equation}
for $k\neq l$
and $\braket{E_k|{\sigma}_{\text{FE}}|E_k}=0$ for all $k$,
where $\varDelta{H}_{\mathrm{eff}}\equiv {H}_{\mathrm{eff}}-H_0=O(\omega^{-1})$,
$p_{\text{can}}^{(k)}=e^{-\beta E_k}/Z$ is the Boltzmann weight,
and $\gamma_{kl}\equiv\sum_{i}(\Gamma_{ik}+\Gamma_{il})/2$
represents the symmetric decay-rate matrix
(see Supplemental Material for the generalization to degenerate $H_0$).
We call ${\sigma}_{\text{FE}}$ the Floquet engineering part
because it describes how the effective Hamiltonian
changes physical observables
from their values in thermal equilibrium.
In contrast to the micromotion part,
the Floquet engineering part contributes to the time-averaged quantities.
As discussed below,
Eq.~\eqref{eq: DB} is regular in the weak dissipation limit $\gamma_{ij}\to0$, where we obtain ${\sigma}_{\text{FE}}$ and hence $\rho_{\mathrm{ness}}(t)$ independent of $\gamma_{ij}$,
and $\rho_{\mathrm{ness}}(t)$ coincides with the canonical Floquet steady state that we define.

Equation~\eqref{eq: FE}
serves as the foundation for the Floquet engineering
in dissipative quantum systems.
Let us imagine, for example,
that an observable ${A}$ has
zero expectation value at thermal equilibrium, $\text{tr}(\rho_\text{can}{A})=0$,
but nonzero value for the NESS, $\text{tr}({\sigma}_{\text{FE}}{A})\neq0$.
This situation means that one can implement 
an appropriate periodic driving $H_{\text{ext}}(t)$ and hence $\varDelta{H}_{\mathrm{eff}}$,
thereby activating the observable ${A}$.
Upon this engineering, Eq.~\eqref{eq: FE}
tells us how much activation is possible for observables of interest.
We will see some examples below.

In addition to their generality,
our formulas [Eqs.~\eqref{eq: MR2}--\eqref{eq: FE}] are extremely efficient
in practical calculations of the nonequilibrium steady state.
In the straightforward calculation,
one numerically integrates the time-dependent Lindblad equation~\eqref{eq: Lindblad}
with a sufficiently small time step
until the system reaches the NESS.
In contrast, 
our formulas enable us to evaluate the NESS
at an arbitrary time $t$ without numerical integration once we have the energy eigenstates $\{\ket{E_k}\}$ of the time-independent Hamiltonian $H_0$.
This difference of efficiency becomes more evident when the Hilbert-space dimension $N$ is large.
In the straightforward calculation, the density matrix $\rho(t)$ is commonly treated
as an $N^2$-dimensional vector and the superoperator $\mathcal{L}_t$ as an $N^2\times N^2$ matrix.
Thus the computational complexity for each time step is $O(N^4)$ in general. On the other hand, the complexity for our formula is one-order smaller and given by  , which derives from the exact diagonalization of $O(N^3)$. Thus our formulas enable us to evaluate the NESS for larger Hilbert-space dimensions that occur, for example, in quantum many-body systems. For special cases where $\mathcal{L}_t$ is a sparse matrix and has only $O(N^0)$ nonzero elements, the computational complexity for one time step is $O(N^2)$. Nevertheless, even for these cases, we need many time steps typically larger than $N$, in obtaining accurate results and, hence, our formulas require less computational complexity.

\section{Numerical verification in a single spin with $S=1$}
By taking a single spin with $S=1$,
we demonstrate how our formula~\eqref{eq: MR2} works
in the quantum dynamics described by the Lindblad equation~\eqref{eq: Lindblad}.
We consider an effective Hamiltonian
for the NV center in diamonds~\cite{Rondin2014}:
\begin{align}\label{eq: HNV}
{H}_\text{NV}(t) &= -B_s S_z + \mathcal{N}_zS_z^2 +\mathcal{N}_{xy}(S_x^2 - S_y^2) +H_{\text{ext}}^{\text{circ}}(t),
\end{align}
where $B_s$ is the static Zeeman field, $\mathcal{N}_z$ and $\mathcal{N}_{xy}$ are 
the coupling constants of magnetic anisotropic terms,
and $H_{\text{ext}}^{\text{circ}}(t)\equiv - B_d(S_x \cos \omega t + S_y \sin \omega t)$
represents the coupling to the circularly-polarized ac magnetic field.
We note that the energy eigenstates of the time-independent part of Eq.~\eqref{eq: HNV} are analytically obtained and our formula can be computed almost analytically in this model.
Since $\mathcal{N}_z\gg \mathcal{N}_{xy}$ in the NV centers~\cite{Rondin2014},
we set $\mathcal{N}_z=1$ and $\mathcal{N}_{xy}=0.05$ in our analysis.

\begin{figure}%[htbp]
\centering
\includegraphics[width = \columnwidth, clip]{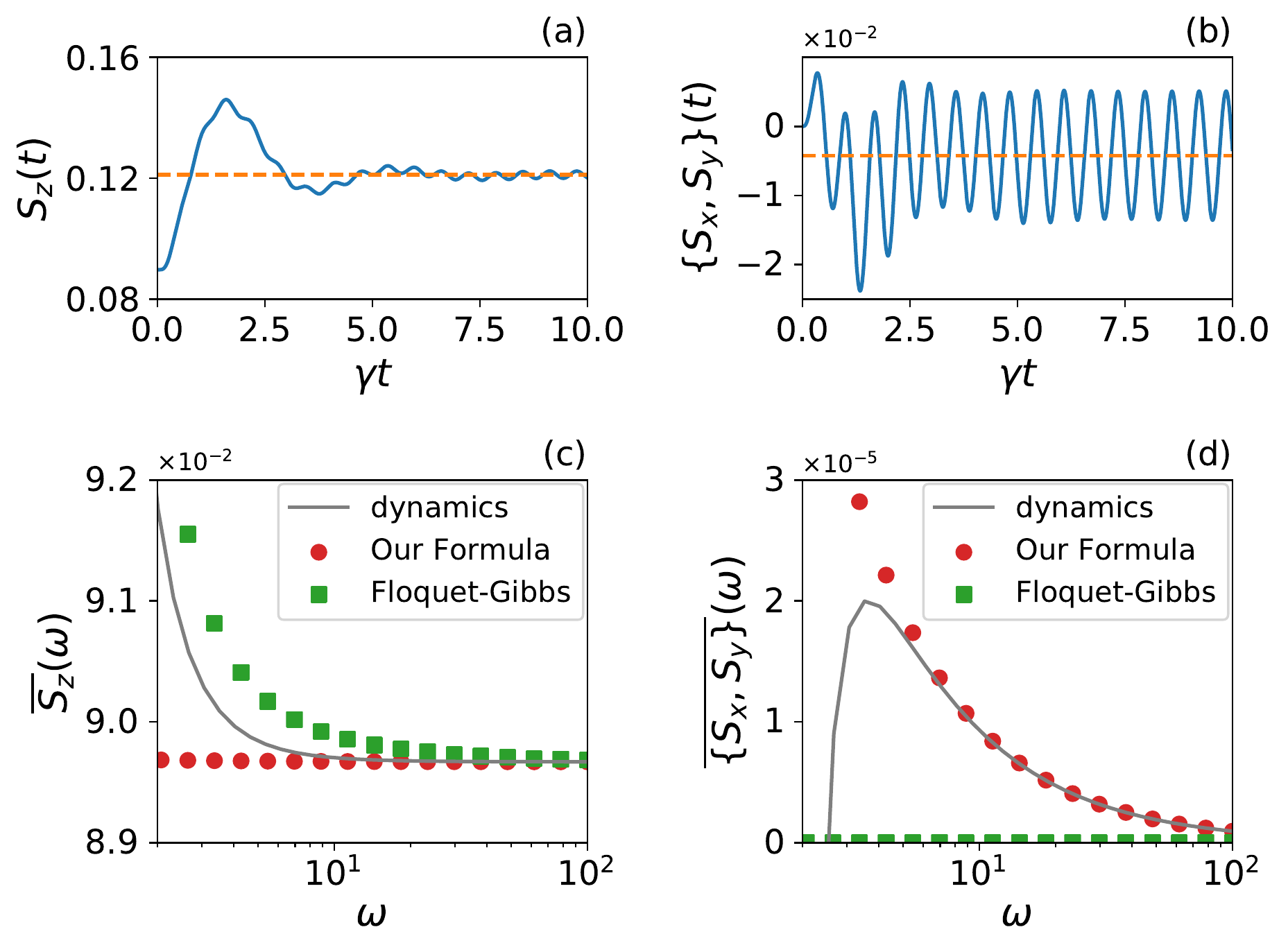}
\caption{(Top) Time evolution of (a) $S_z$ 
and (b) $\{S_x,S_y\}=S_xS_y+S_yS_x$
of a single spin described by the Lindblad equation~\eqref{eq: Lindblad}
with the Hamiltonian~\eqref{eq: HNV}
(see text for the parameters).
The dashed line show the one-cycle average at a sufficiently long time.
(Bottom) One-cycle average $\bar{A}(\omega)$ 
for (c) ${A}=S_z$ and (d) ${A}=\{S_x,S_y\}$
calculated from the dynamics simulation (solid line),
our formula [Eq.~\eqref{eq: MR2}] (circle), and the FGS [Eq.~\eqref{eq: fgs}] (square).
All the other parameters except $\omega$ are the same as in the top panels.
}
\label{fig:approach} %todo unnecessary words in the right
\end{figure} 

Typical time evolutions $O(t)\equiv\mathrm{tr}[\rho(t) {A}]$
are shown for two representative observables ${A}=S_z$ and $\{S_x,S_y\}$ $(=S_xS_y+S_yS_x)$
in Figs.~\ref{fig:approach}(a) and (b), respectively.
In this calculation, we take the thermal state for the time-independent part of ${H}_\text{NV}$ at $t=0$,
and let it evolve according to the Lindblad equation~\eqref{eq: Lindblad} with $H(t)$ being ${H}_\text{NV}(t)$.
The static Zeeman field is $B_s=0.3$,
and the driving parameters are $B_d=0.1$ and $\omega=1$.
As for the Lindblad operators,
we take $\Gamma_{ij}$ according to the heatbath method as
$\Gamma_{ij} = \gamma e^{-\beta E_i}/(e^{-\beta E_i} + e^{-\beta E_j})$
for $i\neq j$ with rate constant $\gamma=0.2$
and $\beta=3$, and $\Gamma_{ii}=0$ for all $i$'s.
Figures~\ref{fig:approach}(a) and (b) show that,
after a sufficiently long time $t \gg \gamma^{-1}$,
the system reaches the nonequilibrium steady state, in which
the observables oscillate with period $T=2\pi/\omega$. 
In particular, the observable ${A}=\{S_x,S_y\}$
is initially zero for a symmetry reason (e.g., $S_x\to-S_x$),
but becomes nonzero on time average.
Namely, this observable is engineered
by the periodic drive $H_{\text{ext}}^{\text{circ}}(t)$.

\begin{figure}
\centering
\includegraphics[width = \columnwidth, clip]{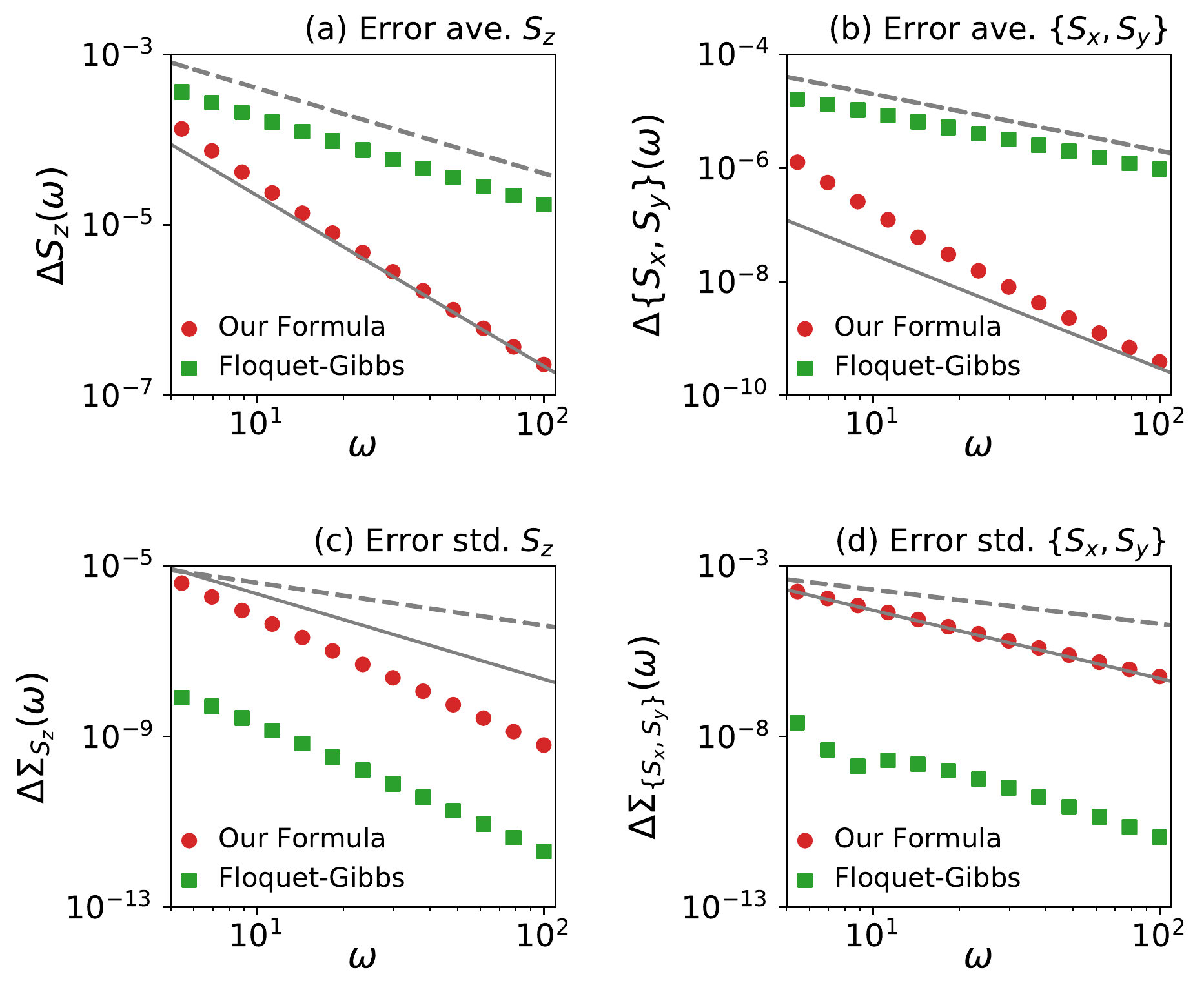}
\caption{
(Top) Difference $\Delta{A}(\omega)$ between the one-cycle average $\bar{A}(\omega)$
calculated from the dynamics simulation and that from our formula [Eq.~\eqref{eq: MR2}] (circle)
or the FGS [Eq.~\eqref{eq: fgs}] (square) plotted against $\omega$ for
(a) ${A}=S_z$ and (b) ${A}=\{S_x,S_y\}$.
(Bottom) Difference $\Delta{\Sigma_A}(\omega)$ between the one-cycle standard deviation $\Sigma_A(\omega)$
calculated from the dynamics simulation and that from our formula [Eq.~\eqref{eq: MR2}] (circle)
or the FGS [Eq.~\eqref{eq: fgs}] (square) plotted against $\omega$ for
(c) ${A}=S_z$ and (d) ${A}=\{S_x,S_y\}$.
In all the panels, the solid and dashed lines are guides for the eye,
showing $\propto\omega^{-2}$ and $\propto\omega^{-1}$, respectively.
}
\label{fig:vs}
\end{figure} 

To test our formula~\eqref{eq: MR2} quantitatively,
we first focus on the one-cycle average
$\bar{A}(\omega)=T^{-1}\int_t^{t+T}ds\,\mathrm{tr}[\rho(s){A}]$
for $t\gg \gamma^{-1}$.
In Figs.~\ref{fig:approach}(c) and (d),
we compare the one-cycle averages calculated
from the actual dynamics
and those calculated from our formulas~\eqref{eq: MR2} and \eqref{eq: FE}
(recall that the micromotion part ${\sigma}_{\text{MM}}(t)$ does not
contribute to the one-cycle averages).
At high frequency $\omega\gtrsim10$,
the difference of the actual dynamics
and our formula decreases quite well.
Defining this difference as $\Delta A(\omega)$,
we plot it against $\omega$
for ${A}=S_z$ and $\{S_x,S_y\}$
in Figs.~\ref{fig:vs}(a) and (b), respectively.
We stress that the difference $\Delta A(\omega)$
decreases more rapidly than $O(\omega^{-1})$
($O(\omega^{-2})$ for $S_z$ and $O(\omega^{-3})$ for $\{S_x,S_y\}$).
This means that our formulas~\eqref{eq: MR2} and \eqref{eq: FE}
perfectly describe the actual NESS at the level of $O(\omega^{-1})$.
As shown in Supplemental Material,
$\Delta A(\omega)=O(\omega^{-2})$ holds true
not only for the two observables but also for all the other observables.
Therefore,
we have verified our formula~\eqref{eq: MR2}
apart from the micromotion part.

For the complementary test of our formula~\eqref{eq: MR2},
we consider the one-cycle standard deviation
$\Sigma_A(\omega)=\{T^{-1}\int_t^{t+T}ds\,\{\mathrm{tr}[\rho(s){A}]-\bar{A}(\omega)\}^2\}^{1/2}$,
which quantifies the micromotion amplitude.
This quantity is suitable for testing our formula~\eqref{eq: MR2}
because it is contributed only by the micromotion part ${\sigma}_{\text{MM}}(t)$.
Since $\Sigma_A(\omega)$ is an $O(\omega^{-1})$ quantity in general,
the accuracy of our formula is verified if the difference $\Delta\Sigma_A(\omega)$ is $O(\omega^{-2})$,
where $\Delta\Sigma_A(\omega)$ is defined by (the absolute value of)
the difference between $\Sigma_A(\omega)$ calculated from the actual dynamics
and that from our formulas~\eqref{eq: MR2} and \eqref{eq: MM}.
This criterion is indeed satisfied
as shown in Figs.~\ref{fig:vs}(c) and (d)
for ${A}=S_z$ and $\{S_x,S_y\}$, respectively.
We remark that our formula leads to $\Sigma_A(\omega)=0$ at $O(\omega^{-1})$ for these observables,
which can be analytically shown by noticing ${H}_{\pm1}\propto{S}_\pm=S_x\pm iS_y$.
Thus the plotted data correspond to $\Sigma_A(\omega)$ itself for the actual dynamics,
and $\Delta\Sigma_A(\omega)$ could be reduced by dealing with
the higher-order terms in Eq.~\eqref{eq: MR2}.
In any case, the fact that $\Delta\Sigma_A(\omega)=O(\omega^{-2})$
justifies our formulas~\eqref{eq: MR2} and \eqref{eq: MM}.

\section{Comparison with the Floquet-Gibbs state}
Let us make comparisons with the Floquet-Gibbs state (FGS),
which has been a candidate for the ensemble description
of the periodically driven dissipative quantum systems~\cite{ShiraiT15, ShiraiT16, LiuD15}.
To define the FGS, we introduce the Floquet state $\ket{u_i(t)}$ 
and its quasienergy $\epsilon_i$.
According to the Floquet theorem,
the time-dependent Schr\"{o}dinger equation
$i\frac{d}{d t}\ket{\psi(t)} =H (t)\ket{\psi(t)}$
has the independent solutions
$\ket{\psi_i(t)}= e^{-i\epsilon_i t}\ket{u_i(t)}$ $(i=1,2,\dots,N)$
with periodicity $\ket{u_i(t+T)}=\ket{u_i(t)}$.
In terms of the Floquet states,
the FGS is defined by
\begin{align}\label{eq: fgs}
    {\rho}_\text{FG} (t) = \frac{1}{Z_\text{FG}}\sum_i e^{-\beta \epsilon_i} \ket{u_i(t)}\bra{u_i(t)},
\end{align}
where $Z_\text{FG}=\sum_i e^{-\beta \epsilon_i}$.
To obtain the Floquet states and quasienergies in practice,
the common method, which we employ here, is to calculate the one-cycle unitary evolution
${U}(T)=\mathbb{T}\exp[-i\int_0^T ds\, H(s)]$,
where $\mathbb{T}$ denotes the time-ordered exponential,
by numerical integrations of the time-dependent Schr\"{o}dinger equation.
The eigenvectors and eigenvalues of ${U}(T)$ correspond to $\ket{u_i(0)}$
and $e^{-i\epsilon_i T}$, which give us $\epsilon_i$ and $\ket{u_i(t)}$.
Note that the Floquet states and quasienergies thus obtained
are exact and involve all-order contributions in $1/\omega$.

Quantitative comparisons
between the actual dynamics and FGS are shown in Fig.~\ref{fig:vs}.
In panel (a) and (d),
we plot the difference of the one-cycle average
calculated from the actual dynamics and FGS
for the two representative observables ${A}=S_z$ and $\{S_x,S_y\}$.
Remarkably, the difference is $O(\omega^{-1})$,
meaning that the FGS cannot reproduce the leading-order
contribution of the one-cycle average~\footnote{
Our results do not contradict the previous studies~\cite{ShiraiT15,ShiraiT16}
presenting a set of sufficient conditions for the FGS being valid
since our example model does not satisfy these conditions.
}
in contrast to our formula~\eqref{eq: MR2}.
As for the micromotion amplitude, or the one-cycle standard deviation $\Sigma_A(\omega)$,
the FGS reproduce the actual values better than our formulas~\eqref{eq: MR2} and \eqref{eq: MM}.
This is partly because the FGS involve all-order contributions in $1/\omega$
while our formulas are the leading-order approximation.
As stated above, our formula could be improved order-by-order
when we start over from our first main result~\eqref{eq: MR1}.

A weak point of the FGS is highlighted in Fig.~\ref{fig:approach}(d),
in which the FGS gives $\overline{\{S_x,S_y\}}(\omega)=0$ for any $\omega$
while it is not true in the actual dynamics.
This is due to an antiunitary dynamical symmetry constraining the Floquet states and hence the FGS.
In fact, we take an antiunitary operator ${V}$:
${V}S_y{V}^\dag=-S_y$ and ${V}{S}_\alpha{V}^\dag={S}_\alpha$ ($\alpha=x$ and $z$).
Then we notice the dynamical symmetry ${V}{H}_\text{NV}(T-t){V}^\dag={H}_\text{NV}(t)$,
which implies that $\ket{\tilde{u}_i(t)}\equiv{V}\ket{u_i(T-t)}$
is also a Floquet state with quasienergy $\epsilon_i$.
Assuming that quasienergies are not degenerate as in our examples,
we have that $\ket{\tilde{u}_i(t)}$ and $\ket{u_i(t)}$ are equivalent
up to an overall phase shift.
Owing to ${V}{A}{V}^\dag=-{A}$ with ${A}=\{S_x,S_y\}$,
the one-cycle averages of ${A}$ calculated for $\ket{\tilde{u}_i(t)}$ and $\ket{u_i(t)}$
differ by their signs, meaning that the one-cycle average vanishes in fact.
Note that similar arguments apply to other observables satisfying ${V}{A}{V}^\dag=-{A}$.

We remark that the dissipation can break such an antiunitary dynamical symmetry
and this is the origin of the nonzero one-cycle average of $\overline{\{S_x,S_y\}}(\omega)$.
We can show that this average vanishes by taking the limit $\gamma_{ij}\to0$ in Eq.~\eqref{eq: FE}.
In other words, the NESS in dissipative systems shows richer properties
inferred only from the effective Hamiltonian itself.
Our formula~\eqref{eq: MR2} well describes these properties
unlike the FGS~\eqref{eq: fgs}, which incorporates
no information about the dissipation, or the Lindblad operators.

One might be interested in an approximate description
of the NESS independent of the details of $\gamma_{ij}$ for weak dissipation,
and expect that the FGS serves as such a description.
Interestingly, this is not true
at least within our formulation of periodically driven dissipative
systems described by Eqs.~\eqref{eq: Lindblad} and \eqref{eq: DB}.
Instead, the actual NESS coincides with yet another state which we name the canonical Floquet steady state (CFSS)
defined by replacing the quasienergy $\epsilon_i$ by the real energy $E_i$ in Eq.~\eqref{eq: fgs}.
One can show this by comparing the high-frequency expansion of the CFSS
and our formulas [Eqs.~\eqref{eq: MR1}--\eqref{eq: FE}] in the limit of $\gamma_{ij}\to0$
(see Supplemental Material for details).

\section{Discussions and Conclusions}
We have derived and verified
the simple and general formulas [Eqs.~\eqref{eq: MR1}--\eqref{eq: FE}]
describing the NESS in dissipative quantum systems
under high-frequency periodic drivings.
We have also exemplified the dynamical symmetry breaking
and the possibility of the Floquet engineering in driven dissipative systems
in the NV centers in diamonds.
Being quite general,
our formulas would play the fundamental role in understanding and engineering
unusual nonequilibrium states in various quantum systems
such as atoms and molecules, trapped ions, condensed matters, and so on.

\begin{figure}
\centering
\includegraphics[width = 6cm, clip]{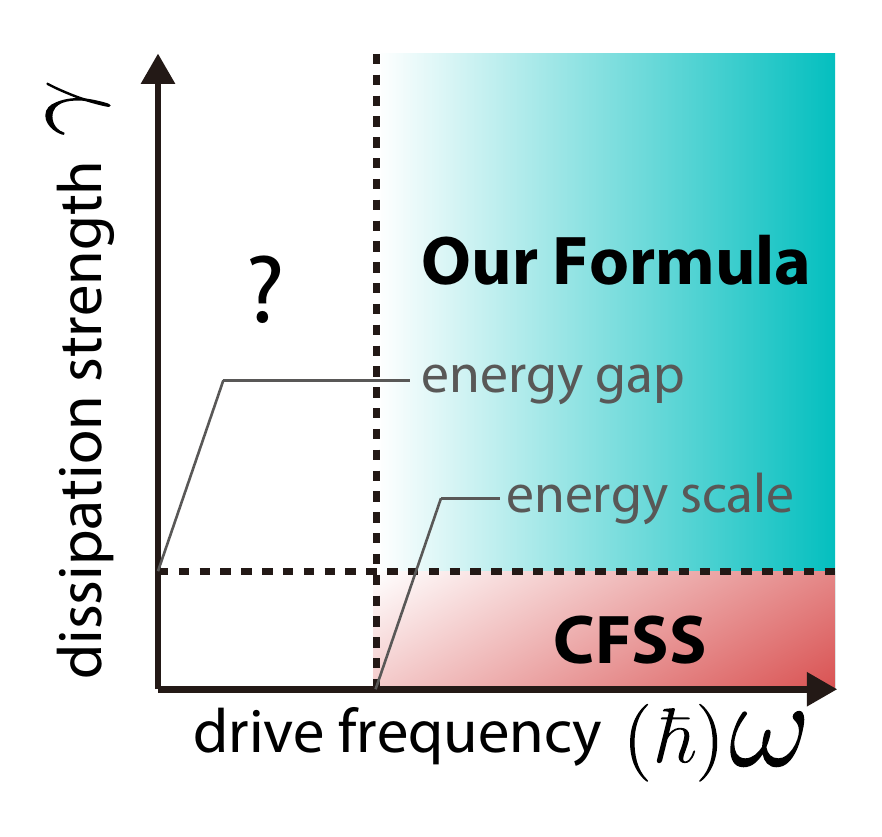}
\caption{
Validity of NESS formulas in parameter space. Our formula is valid when the driving frequency $\omega$ is larger than energy scales of the system. When the dissipation strength is smaller than the (nonzero) minimum energy gap, our formula reduces to the CFSS. At lower frequency, the NESS description remains an open question.
}
\label{fig:region}
\end{figure}

The parameter region in which our formulas are valid is depicted in Fig.~\ref{fig:region}. Based on the high-frequency expansion, our formulas are valid when the driving frequency $\omega$ (more precisely, the photon energy $\hbar\omega$) is greater than the energy scales of the system and the system-drive coupling. We note, however, that our formulas hold true for any strength of dissipation $\Gamma_{ij}$ (or $\gamma_{ij}$) within the Floquet-Lindblad equation~\eqref{eq: Lindblad}. As we have shown, our formulas reduce to the CFSS rather than the FGS when the dissipation strength is smaller than the energy gap, i.e., the nonzero minimum difference between eigenenergies (our formulas are generalized for the degenerate Hamiltonian in Supplementary Materials).

One should note that the Floquet-Lindblad equation~\eqref{eq: Lindblad} becomes invalid when $\Gamma_{ij}$ is too large. The Lindblad-type dissipation is derived from several approximations such as the Born-Markov approximation~\cite{BreuerH02}. These approximations require the condition that the relaxation time $\sim1/\Gamma_{ij}$ is longer than the time scale of system's dynamics and the bath correlation time. Namely, the dissipation rate $\Gamma_{ij}$ should be smaller than the other relevant energy scales. Note that the high-frequency driving does not break this condition while lower frequencies may be problematic.

We remark a further generalization of our results Eqs.~\eqref{eq: MR1}--\eqref{eq: FE}. Although we have assumed the detailed balance condition~\eqref{eq: DB}, this condition can be removed as long as the transition-rate matrix $\Gamma_{ij}$ is irreducible. In this case, the solution of the Lindblad equation without driving is not the canonical ensemble $\rhocan$ but another state $\tilde{\rho}$ characterized by $-i[H_0,\tilde{\rho}]+\mD(\tilde{\rho})=0$. Correspondingly, our results for the NESS [Eqs.~\eqref{eq: MR1}--\eqref{eq: FE}] hold true with the following replacements: $\rhocan\to\tilde{\rho}$ and $p_\text{can}^{(k)}\to\tilde{p}_\text{can}^{(k)}=\braket{E_k|\tilde{\rho}|E_k}$. Thus our formulas apply to any periodically driven dissipative systems as long as the dissipation is of Lindblad type and irreducible. Therefore, our formulas are useful for a broad class of systems in exploring generic features of the NESS and in estimating Floquet-engineered physical quantities. 

It remains an open question to find a simple and general formula for the NESS at lower frequency. The applicability of the CFSS in many-body systems is also a nontrivial issue because the energy gap can be very small in those systems. Addressing these issues will lead us to the complete understanding of the NESS in dissipative Floquet systems.

\section*{Acknowledgements}
The authors are grateful to
Sho Higashikawa and Hiroyuki Fujita for collaboration
on the early stage of this work,
and to
Koki Chinzei,
Takashi Mori, Takahiro Sagawa, Tatsuhiko Shirai,
and Hirokazu Tsunetsugu
for fruitful discussions.
T.N.I. was supported by JSPS KAKENHI Grant No. JP18K13495.
M.S. was supported by Grant-in-Aid for Scientific Research on Innovative Area,
``Physical Properties of Quantum Liquid Crystals'' (Grant No. 19H05825)
and JSPS KAKENHI Grant No. JP17K05513 and JP20H01830.

%\bibliography{../../../../Bibtex/cfss,DD_bib_higashi}

%\newpage
%\phantom{a}
%\newpage
%%%%%%%%%%%%%%%%%%%%%%%%%%%%%%%%%%%%%%%%
\setcounter{figure}{0}
\setcounter{equation}{0}
\setcounter{section}{0}

\onecolumngrid

\begin{center}

\vspace{1.5cm}

{\large \bf Supplemental Material:
General description for nonequilibrium steady states\\in periodically driven dissipative quantum systems}

\vspace{0.3cm}

{\large Tatsuhiko N. Ikeda$^{1}$ and Masahiro Sato$^{2}$} \\[2mm]
$^1$\textit{The Institute for Solid State Physics, The University of Tokyo, Kashiwa, Chiba 277-8581, Japan}\\
$^2$\textit{Department of Physics, Ibaraki University, 
Mito, Ibaraki 310-8512, Japan}

\end{center}

\vspace{0.6cm}

\renewcommand{\theequation}{S\arabic{equation}}
\renewcommand{\thesection}{S\arabic{section}}
\renewcommand{\thefigure}{S\arabic{figure}}
%\makeatletter
% \renewcommand{\@cite}[2]{[S#1]}
% \renewcommand{\@biblabel}[1]{[S#1]}
%\makeatother

%todo tilted magnetic field \to slanted magnetic field
%todo rotating wave frame \to rotating frame
%todo a slight deviation can be attributed to @@@

\section{High-frequency expansion for the Lindblad equation}\label{app:HFexpansion}
The high-frequency expansion has been developed
in unitary dynamics
and there are several formulations as summarized in Ref.~\cite{MikamiT16}.
For the Lindblad equation,
the high-frequency expansion has been discussed
in terms of the Floquet-Magnus formalism in Ref.~\cite{DaiC16}.
In this paper,
we make use of the high-frequency expansion of the Lindblad equation
in terms of the van Vleck approach,
which we describe below for completeness.

%########
The Lindblad equation that we discuss in this work
is symbolically represented as
\begin{align}\label{seq:lindt}
	\partial_t \hrho(t) = \mL(t) \hrho(t),
\end{align}
where the time-dependent Liouvillian $\mL(t)$ is defined by
\begin{align}
	\mL(t) \hrho = -i [\htot(t),\hrho] +\mD(\hrho),
\end{align}
where $\htot(t)=\htot(t+T)$ is the periodic Hamiltonian and $\mD(\hrho)$ denotes the dissipation term represented by the Lindblad operators.
We introduce the Fourier series for the Liouvillian as
\begin{align}
	\mL(t) = \sum_m \mL_m e^{-im\omega t}.
\end{align}
Since the Lindblad operators $\lind_{ij}$ are time-independent in this work,
each Fourier component is given as follows:
\begin{align}
	\mL_0\hrho = -i[\hH_0,\hrho] +\mD(\hrho);\qquad
	\mL_m\hrho = -i[\hH_m,\hrho] \quad (m\neq0).\label{seq:Lm}
\end{align}

%#########
The formal solution of Eq.~\eqref{seq:lindt}
is obtained as $\hrho(t)=\mV(t,t')\hrho(t')$
with the propagator
\begin{align}
	\mV(t,t') = \mathbb{T}\exp\left[ \int_{t'}^t \mL(s) ds\right],
\end{align}
where $\mathbb{T}\exp$ denotes the time-ordered exponential.
The determining equations for $\mV$ are
\begin{align}
	\partial_t \mV(t,t') &= \mL(t) \mV(t,t'),\label{seq:det1}\\
	\mV(t',t') &= 1.\label{seq:det2}
\end{align}

%######
The high-frequency expansion in terms of the van Vleck approach
makes the following ansatz:
\begin{align}\label{seq:ansatz}
	\mV(t,t') = e^{\mG(t)}e^{(t-t')\Leff}e^{-\mG(t')},
\end{align}
where $\mG(t)$ is periodic in time and $\Leff$ is time-independent.
This ansatz satisfies Eq.~\eqref{seq:det2} for any choices
of $\mG(t)$ and $\Leff$,
and what determines these two is Eq.~\eqref{seq:det1}.
As we will see below, Eq.~\eqref{seq:det1} only determines
the derivative of $\mG(t)$
and thus we further impose $\int_0^T \mG(t)dt=0$
to fix the constant of integration.

%####
To obtain the determining equations for $\mG(t)$ and $\Leff$,
we substitute Eq.~\eqref{seq:ansatz} into Eq.~\eqref{seq:det1},
having
\begin{align}\label{seq:deransatz1}
	\partial_t (e^{\mG(t)}) +  e^{\mG(t)}\Leff = \mL(t)e^{\mG(t)}.
\end{align}
To rewrite the first term on the left-hand side
(see Ref.~\cite{Mananga2011} for the case of unitary dynamics),
we invoke the Wilcox formula~\cite{Wilcox1967}
$\partial_\lambda e^{-\beta H}=-\int_0^\beta e^{-uH}(\partial_\lambda H)e^{-(\beta-u)H}du$
for $H=H(\lambda)$.
We replace $\lambda$, $H$, and $\beta$
by $t$, $-\mG$, and $1$, respectively,
obtaining
\begin{align}
	\partial_t(e^{\mG(t))})
	=\left\{\int_0^1 e^{u\mG(t)}[\partial_t\mG(t)]e^{-u\mG(t)}du\right\}e^{\mG(t)}
	=\left\{\int_0^1 e^{u\ad_\mG}du
	[\partial_t\mG(t)]\right\}e^{\mG(t)}
	=\{\phi(\ad_\mG)[\partial_t\mG(t)]\}e^{\mG(t)}.\label{seq:derG}
\end{align}
Here $\ad_\mG$ is defined by $\ad_\mG\rho=[\mG(t),\rho]$
and $\phi(x)\equiv (e^x-1)/x$.
We substitute Eq.~\eqref{seq:derG} into Eq.~\eqref{seq:deransatz1}
and have
\begin{align}
	\partial_t \mG(t) = \phi^{-1}(\ad_\mG)\mL(t)-\phi^{-1}(\ad_\mG)e^{\ad_\mG}\Leff.
\end{align}
Now we notice $\phi^{-1}(x)e^x = \phi^{-1}(-x)$
and make use of the Taylor expansion of $\phi^{-1}(x)$: $\phi^{-1}(x)=\sum_{k=0}^\infty \frac{B_k}{k!}x^k$,
where $B_k$ denotes the Bernoulli number
($B_0=1$, $B_1=-1/2$, $B_2=1/6\cdots$).
Then we obtain
\begin{align}\label{seq:deteq}
	\partial_t\mG(t) = \sum_{k=0}^\infty \frac{B_k}{k!}(\ad_\mG)^k \left[ \mL(t)+(-1)^{k+1}\Leff \right].
\end{align}

%######
Now we determine $\mG(t)$ and $\Leff$
from Eq.~\eqref{seq:deteq}
by the series expansions
\begin{align}
	\mG(t) = \sum_{k=1}^\infty \mG^{(k)}(t);
	\qquad
	\Leff = \sum_{k=1}^\infty \Leff^{(k)}.
\end{align}
We substitute these expansions into Eq.~\eqref{seq:deteq}
and find the order-by-order solutions,
where we assign an order 1 for $\mL(t)$
and $k$ for $\mG^{(k)}(t)$ and $\Leff^{(k)}$
(see Ref.~\cite{MikamiT16} for the case of unitary dynamics).

%###
The first-order equation leads to
\begin{align}\label{seq:1storder}
	\partial_t \mG^{(1)}(t)
	=\mL(t)-\Leff^{(1)}.
\end{align}
To obtain $\Leff^{(1)}$,
we integrate Eq.~\eqref{seq:1storder}
over $0\le t\le T$.
With the periodicity $\mG(T)=\mG(0)$,
we obtain
\begin{align}\label{seq:Leff1}
	\Leff^{(1)} = \int_0^T \frac{dt}{T}\mL(t)
	=\mL_0.
\end{align}
To obtain $\mG(t)$,
we integrate Eq.~\eqref{seq:1storder},
having
\begin{align}
	\mG^{(1)}(t)-\mG^{(1)}(0)=\int_0^t \mL(s)ds-t\Leff^{(1)}
	=t\mL_0 +\sum_{m\neq0}\frac{e^{-im\omega t}-1}{-im\omega}\mL_m -t\Leff^{(1)},
\end{align}
which means
\begin{align}
	\mG^{(1)}(t) = \frac{i}{\omega}\sum_{m\neq0}\frac{e^{-im\omega t}}{m}\mL_m.
\end{align}
Note that $\Leff^{(1)}$ is $O(\omega^0)$
and $\mG^{(1)}(t)$ is $\oo$.

%####
The second-order equation leads to
\begin{align}\label{seq:2ndorder}
	\partial_t\mG^{(2)}(t)=-\frac{1}{2}[\mG^{(1)}(t),\mL(t)+\Leff^{(1)}] - \Leff^{(2)}.
\end{align}
To obtain $\Leff^{(2)}$,
we integrate Eq.~\eqref{seq:2ndorder}
over $0\le t\le T$.
Upon this, we note that $\mG^{(2)}(t)$ is periodic and $\int_0^T\mG^{(1)}(t)dt=0$.
Then we have
\begin{align}
	\Leff^{(2)} = -\frac{1}{2}\int_0^T\frac{dt}{T}[\mG^{(1)}(t),\mL(t)]
	=-\frac{i}{2\omega}\sum_{m\neq0}\frac{[\mL_m,\mL_{-m}]}{m}
	=-\frac{i}{\omega}\sum_{m>0}\frac{[\mL_m,\mL_{-m}]}{m}.\label{seq:Leff2}
\end{align}
By straightforward calculations,
one can obtain $\mG^{(2)}(t)$
by integrating Eq.~\eqref{seq:2ndorder}
from $0$ to $t$.
Likewise,
one could systematically build
the higher order solutions
although we do not go further here.

%####
Let us rewrite $\Leff=\Leff^{(1)}+\Leff^{(2)}$
in terms of the effective Hamiltonian $\Heff$~\cite{EckardtA15,MikamiT16}:
\begin{align}\label{seq:Heff}
	\Heff = \hH_0 + \frac{1}{\omega}\sum_{m>0}\frac{[\hH_{-m},\hH_{m}]}{m}+\ot.
\end{align}
To do this, we consider the action of $\Leff^{(2)}$ onto a density operator $\hrho$.
From Eqs.~\eqref{seq:Leff2}
and \eqref{seq:1storder},
we have
\begin{align}
	\Leff^{(2)}\hrho
	&=-\frac{i}{\omega}\sum_{m>0}\frac{1}{m}
	(\mL_m\mL_{-m}\hrho - \mL_{-m}\mL_m \hrho)
	=\frac{i}{\omega}\sum_{m>0}\frac{1}{m}
	([\hH_m,[\hH_{-m},\hrho]] - [\hH_{-m},[\hH_m, \hrho]) \nonumber\\
	&=-\frac{i}{\omega}\sum_{m>0}\frac{1}{m}
	[[\hH_{-m},\hH_{m}],\hrho],\label{seq:Leff2}
\end{align}
where we have used the Jacobi identity
$[A,[B,C]]+[B,[C,A]]+[C,[A,B]]=0$.
Combining Eqs.~\eqref{seq:Lm}, \eqref{seq:Leff1}, \eqref{seq:Leff2},
and \eqref{seq:Heff},
we obtain
\begin{align}\label{app:Leff}
	\Leff\hrho = -i[\Heff,\hrho] +\mD(\hrho) + \ot.
\end{align}
We remark that $\Leff\hrho$ is not equal to $-i[\Heff,\hrho] +\mD(\hrho)$
at higher orders
because $\Leff$ involves contributions of $\mD$
from $\ot$.

%##############################################################################
\section{Derivation of the main result [Eqs.~(8)--(10)]}\label{app:derivationMR}
Here we derive Eqs.~(8)--(10) from Eq.~(7) in the main text.
For this purpose,
we solve $\Leff\rhoinf=0$ for $\rhoinf$
at the leading order of $\omega^{-1}$.
It is convenient to work
in the energy eigenbasis
and separate the diagonal and off-diagonal parts:
\begin{align}
 \rhoinf &=\rhoinfd+\rhoinfoff,\\
 \rhoinfd &=\sum_k \rhoinfel{kk}\ket{E_k}\bra{E_k},\\
 \rhoinfoff &=\sum_{k,l (k\neq l)} \rhoinfel{kl}\ket{E_k}\bra{E_l},
\end{align}
where $\ket{E_k}$ denotes the eigenstate of $\hzero$
with eigenenergy $E_k$.

%####
First, we consider the off-diagonal elements of
both sides of $\Leff\rhoinf=0$,
having, for $k\neq l$,
\begin{align}\label{eq:offel}
\braket{E_k|\Leff\rhoinf|E_l}
=[-i(E_k-E_l)-\gamma_{kl}]\rhoinfel{kl}
-i(\rhoinfel{ll}-\rhoinfel{kk})\braket{E_k|\dHeff|E_l}
-i\braket{E_k|[\dHeff,\rhoinfoff]|E_l}=0,
\end{align}
where $\gamma_{kl}\equiv\sum_{i}(\Gamma_{ik}+\Gamma_{il})/2$
and $\dHeff\equiv \Heff-\hzero=O(\omega^{-1})$.
Equation~\eqref{eq:offel} is transformed as 
\begin{align}\label{eq:offelrec}
\rhoinfel{kl}=\frac{\braket{E_k|\dHeff|E_l}}{(E_k-E_l)-\gamma_{kl}}(\rhoinfel{kk}-\rhoinfel{ll})
-\frac{\braket{E_k| [\dHeff,\rhoinfoff] |E_l}}{(E_k-E_l)-i\gamma_{kl}}.
\end{align}
Note that the denominators $(E_k-E_l)-i\gamma_{kl}$ do not vanish
since $\gamma_{kl}>0$ is ensured by the nonnegativity and irreducibility of $\Gamma_{ij}$.
Now, as a working hypothesis, we suppose that the diagonal elements $\rhoinfel{kk}$ are $O(\omega^0)$
as verified later.
Then the first term on the right-hand side of Eq.~\eqref{eq:offelrec} is $O(\omega^{-1})$
since $\dHeff=O(\omega^{-1})$.
Notice that the second term depends only on the off-diagonal elements $\rhoinfel{kl}$
and Eq.~\eqref{eq:offelrec} can be solved recursively.
This yields the $\omega^{-1}$ expansion for the off-diagonal elements $\rhoinfel{kl}$,
whose leading order contribution is given by
\begin{align}\label{eq:offellead}
\rhoinfel{kl}=\frac{\braket{E_k|\dHeff|E_l}}{(E_k-E_l)-i\gamma_{kl}}(\rhoinfel{kk}-\rhoinfel{ll})+O(\omega^{-2}).
\end{align}

%####
Next, we consider the diagonal elements of both sides of $\Leff\rhoinf=0$,
having
\begin{align}\label{eq:del}
	\braket{E_k|\Leff\rhoinf|E_k}
	=-i \braket{E_k| [\dHeff,\rhoinfoff] |E_l}
	+\sum_l( \Gamma_{kl}\rhoinfel{ll} - \Gamma_{lk}\rhoinfel{kk})=0.
\end{align}
We note $\braket{E_k| [\dHeff,\rhoinfoff] |E_l}=O(\omega^{-2})$
since both $\dHeff$ and $\rhoinfoff$ are $O(\omega^{-1})$.
Thus the diagonal elements $\rhoinfel{kk}$ are determined up to $O(\omega^{-1})$
by the equation:
\begin{align}
	\sum_{l}( \Gamma_{kl}\rhoinfel{ll} - \Gamma_{lk}\rhoinfel{kk} )=0.
\end{align}
According to the irreducibility and the detailed balance condition of $\Gamma_{kl}$,
we have the unique solution as
\begin{align}\label{eq:delfin}
	\rhoinfel{kk} = \canw^{(k)} = \frac{e^{-\beta E_k}}{Z},	
\end{align}
where the error is $O(\omega^{-2})$.
This result means
\begin{align}
	\rhoinfd = \rhocan + O(\omega^{-2}).
\end{align}
Since these diagonal elements $\rhoinfel{kk}$ are $O(\omega^0)$,
the working hypothesis introduced above has been verified.
By substituting Eq.~\eqref{eq:delfin} into Eq.~\eqref{eq:offellead},
we have the leading-order expression for the off-diagonal elements:
\begin{align}
	\rhoinfel{kl}=\frac{\braket{E_k|\dHeff|E_l}}{(E_k-E_l)-i\gamma_{kl}}(\canw^{(k)}-\canw^{(l)})+O(\omega^{-2})= \braket{E_k|\sigfe|E_l} +O(\omega^{-2}),
\end{align}
which implies
\begin{align}
	\rhoinfoff = \sigfe + O(\omega^{-2}).
\end{align}
We remark that $\text{tr}(\sigfe)=0$ since the each of the diagonal elements of $\sigfe$ vanishes.

%#####
Now that we have obtained the leading-order expression for $\rhoinf=\rhoinfd+\rhoinfoff$,
let us calculate the time-dependent density matrix
by $\hrho(t)=e^{\mG(t)}\rhoinf$.
By noticing that $\rhoinfd$ is $O(\omega^0)$ and $\rhoinfoff$ is $O(\omega^{-1})$
and using the Taylor expansion $e^{\mG(t)}=1+\mG(t)+O(\omega^{-2})$,
we obtain
\begin{align}
	\hrho(t) &= \rhocan +\mG(t)[\rhoinfd] + \sigfe + O(\omega^{-2})  \nonumber\\
	&= \rhocan + \sigmm(t) + \sigfe + O(\omega^{-2}),\label{eq:MR}
\end{align}
where we have defined
\begin{align}
	\sigmm(t) = \mG(t)[\rhoinfd]
	= \frac{1}{\omega}\sum_{m\neq0}\frac{e^{-im\omega t}}{m}[H_m,\rhocan].
\end{align}
Thus we have derived the Eqs.~(8)--(10) in the paper.
We remark $\text{tr}[\sigmm(t)]=0$, which follows from the cyclic property of trace,
and hence $\text{tr}[\hrho(t)]=1$, at least, up to this order.

%##########################################################################################
\section{Generalization to the degenerate energy spectra}
Here we generalize our main results [Eqs.~(8)--(10)]
to the cases in which the energy spectrum $\{E_i\}_{i=1}^N$ is degenerate.
To deal with such a spectrum, we introduce new notations
for the eigenenergies and eigenstates of $\hzero$
given by $E_i^\alpha$ and $\ket{E_i^\alpha}$ with $\hzero\ket{E_i^\alpha}=E_i\ket{E_i^\alpha}$.
Here, $i$ $(=1,\dots,M)$ labels the distinct eigenenergies
and $\alpha$ $(=1,2,\dots,N_i)$ does the degenerate eigenstates,
and we assume the orthonormality $\braket{E_i^\alpha|E_j^{\alpha'}}=\delta_{ij}\delta_{\alpha\alpha'}$.
We remark that the choice of the degenerate eigenstates
has arbitrariness up to unitary transformation
for each degenerate subspace:
\begin{align}\label{app:unitary}
	\ket{E_i^\alpha} \to \tket{E_i^\alpha} =\sum_{\beta=1}^{N_i} \ket{E_i^\beta}U_{\beta\alpha}^{(i)},
\end{align}
where $U^{(i)}$ is an $N_i\times N_i$ unitary matrix.
We should be aware that the following formulation
needs to be invariant under the unitary transformation~\eqref{app:unitary}.

%####
The Lindblad operators with the detailed balance condition
are generalized as follows: $\lind_{ij}\to\lind_{i\alpha,j\alpha'}\equiv\ket{E_i^\alpha}\bra{E_j^{\alpha'}}$.
The corresponding transition rates
are written as $\Gamma_{i\alpha,j\alpha'}$,
which are assumed independent of the degeneracy
labels $\alpha$ or $\alpha'$
and to 
%These rates are assume to 
satisfy the detailed balance condition:
\begin{align}\label{app:DB}
\Gamma_{i\alpha,j\alpha'}\Exp{-\beta E_j} = \Gamma_{j\alpha',i\alpha}\Exp{-\beta E_i} \qquad (\text{for}\ i\neq j),
\end{align}
and $\Gamma_{i\alpha,j\alpha'}=0$ for $i=j$.
We also assume that the transition rates $\Gamma_{i\alpha,j\alpha'}$
are irreducible, which is satisfied, for example, if
$\Gamma_{i\alpha,j\alpha'}>0$ for all pairs of $i$ and $j$.
Then the dissipation term in the Lindblad equation
is given by
\begin{align}\label{app:dterm}
	\mD(\hrho)=\sum_{\substack{i\alpha,j\beta\\ (i\neq j)}}
	\Gamma_{i\alpha,j\beta}
	\left( \lind_{i\alpha,j\beta}\hrho\lind^\dag_{i\alpha,j\beta}
	-\frac{1}{2}\{ \lind_{i\alpha,j\beta}^\dag \lind_{i\alpha,j\beta},\hrho\}\right).
\end{align}
As one can check easily,
the dissipation term~\eqref{app:dterm}
is invariant under Eq.~\eqref{app:unitary}.

Now that we have the Lindblad equation,
we can repeat the high-frequency-expansion arguments in Sec.~\ref{app:HFexpansion}
to obtain Eq.~\eqref{app:Leff} for the generalized $\mD$ term~\eqref{app:dterm}.
Thus, we move on to deriving the counterparts of the main results [Eqs.~(8)--(10)]
by generalizing the arguments in Sec.~\ref{app:derivationMR}.

%###
Let us solve $\Leff\rhoinf=0$ for $\rhoinf$
at the leading order of $\omega^{-1}$.
The solution $\rhoinf$ is necessarily
written in the following form:
\begin{align}
\rhoinf &=\rhoinfd+\rhoinfoff,\\
\rhoinfd &=\sum_{k,\alpha\beta} \rhoinfel{k\alpha,k\beta}\ket{E_k^\alpha}\bra{E_k^\beta},\\
\rhoinfoff &=\sum_{\substack{k\alpha,l\beta \\(k\neq l)}} \rhoinfel{k\alpha,l\beta}\ket{E_k^\alpha}\bra{E_l^\beta}.
\end{align}
Since we have arbitrariness
of choosing the degenerate eigenstates
as noted above,
we can assume
without loss of generality that $\rhoinfd$
is diagonal
\begin{align}
\rhoinfel{k\alpha,k\beta}=q_{k\alpha} \delta_{\alpha\beta},
\end{align}
where $q_{k\alpha}\ge0$.

%###
First, we focus on the off-diagonal matrix elements of $\Leff\rhoinf=0$:
$\braket{E_k^\alpha|\Leff\rhoinf|E_l^\beta}=0$.
Repeating similar arguments in deriving Eq.~\eqref{eq:offellead},
we have
\begin{align}\label{eq:deg:offellead}
\rhoinfel{k\alpha,l\beta}=\frac{\braket{E_k^\alpha|\dHeff|E_l^\beta}}{(E_k-E_l)-i\gamma_{kl}}(q_{k\alpha}-q_{l\beta})+O(\omega^{-2}),
\end{align}
where
we have introduced the working hypothesis $q_{k\alpha}=O(\omega^0)$
and $\gamma_{kl}\equiv\sum_{i,\gamma}(\Gamma_{i\gamma,k\alpha}+\Gamma_{i\gamma,l\beta})/2$ (Remember that $\Gamma_{k\alpha,l\beta}$ is independent of $\alpha$ or $\beta$).

Next, we consider the diagonal elements of $\Leff\rhoinf=0$:
$\braket{E_k^\alpha|\Leff\rhoinf|E_k^\beta}=0$.
While, for $\alpha\neq\beta$, we have irrelevant equations
of $O(\omega^{-2})$,
for $\alpha=\beta$, we have
\begin{align}
	\sum_{l,\beta}( \Gamma_{k\alpha,l\beta}q_{l\beta} - \Gamma_{l\beta,k\alpha}q_{k\alpha} )=0.
\end{align}
According to the irreducibility of $\Gamma_{k\alpha,l\beta}$,
this equation has the unique positive solution,
which is given by
\begin{align}\label{eq:deg:delfin}
q_{k\alpha}=\canw^{(k)} = \frac{e^{-\beta E_k}}{Z},	
%	e^{-\beta E_k}/Z
%	\rhoinfel{kk} = \canw^{(k)} = \frac{e^{-\beta E_k}}{Z},	
\end{align}
with $Z=\sum_{k,\alpha}e^{-\beta E_{k}}$.
One can confirm this by using the detailed balance condition.
From the above argument, we obtain 
\begin{align}
	\rhoinf = \rhocan + \sigfe +O(\omega^{-2}),
\end{align}
where
\begin{align}\label{eq:deg:fe}
	\braket{E_k^\alpha|\sigfe|E_l^\beta} =
	\frac{\braket{E_k^\alpha|\dHeff|E_l^\beta}}{(E_k-E_l)-i\gamma_{kl}}(\canw^{(k)}-\canw^{(l)}) \qquad (k\neq l)
\end{align}
and $\braket{E_k^\alpha|\sigfe|E_l^\beta}=0$ for $k=l$.

Finally, we calculate the time-dependent density matrix
by $\hrho(t)=e^{\mG(t)}\rhoinf$,
obtaining
\begin{align}
	\hrho(t) 
	&= \rhocan + \sigmm(t) + \sigfe + O(\omega^{-2}),\label{eq:deg:MR}
\end{align}
where $\sigmm(t)$ is the same as Eq.~\eqref{eq:MR} for the nondegenerate case.

To summarize, our main results [Eqs.~(8)--(10)] are generalized in a straightforward manner.
Among the three terms on the right-hand side of Eq.~\eqref{eq:deg:MR},
the first two $\rhocan$ and $\sigmm(t)$ are expressed exactly in the same way
for the degenerate case,
and the third one $\sigfe$ is naturally generalized as in Eq.~\eqref{eq:deg:fe}.

%##########################################################################################
\section{All the observables in the effective model for the NV center}
Although we have discussed the two observables $\hSz$ and $\{\hSx,\hSy\}$,
there are in total 8 observables including these two
(since we are considering a spin-1 system represented by $3\times3$ matrices):
the spins along one direction, $\hSx$, $\hSy$ and $\hSz$,
and the nematics $\hSz^2$, $\hSx^2 - \hSy^2$, $\hSx\hSy+\hSy\hSx$,
$\hSy\hSz+\hSz\hSx$, and $\hSz\hSx+\hSx\hSz$.
In this section, we consider all these observables and validate our main results [Eqs.~(8)--(10)].

%#########################
\subsection{Vanishing one-cycle averages due to dynamical symmetry}\label{sec:vanishingave}
We compare the one-cycle average $\bar{A}(\omega)$
of an observable $\hO$ for the actual dynamics
with that from our formulas [Eqs.~(8)--(10)] and the FGS.
Upon this comparison, we note that the average vanishes for 
%all of these vanish for the following observables:
$\hO=\hSx$, $\hSy$, $\hSy\hSz+\hSz\hSx$, and $\hSz\hSx+\hSx\hSz$.
The common property shared by these observables is that
they are all odd under the $\pi$-rotation around the $\hSz$ axis:
\begin{align}\label{eq:obsUodd}
\hUz_\pi \hO \hUzd_\pi=-\hO \qquad \text{for}\ \hO=\hSx,\hSy,\hSy\hSz+\hSz\hSx,\hSz\hSx+\hSx\hSz	
\end{align}
Another important property is the dynamical symmetry associated with this unitary operation:
\begin{align}\label{eq:DS1}
	\hUz_\pi\hnv(t+T/2)\hUzd_\pi=\hnv(t).
\end{align}
As we see below, Eqs.~\eqref{eq:obsUodd} and \eqref{eq:DS1} imply that
the one-cycle averages for these observables vanish in the actual calculation,
our formulas [Eqs.~(8)--(10)], and the FGS, respectively.

%######################
First, we discuss the actual dynamics governed by the Lindblad equation:
\begin{align}\label{eq:lindorg}
\partial_t \hrho(t)=-i[H(t),\hrho(t)]+\mD[\hrho(t)].	
\end{align}
We try to have some implication of the dynamical symmetry~\eqref{eq:DS1} to this equation.
For this purpose, we shift $t\to t+T/2$ in the equation and apply $\hUz_\pi$ from left and $\hUzd_\pi$ from right
to the both sides of the equation,
having $\partial_t\rhoUzp(t) =-i[\hnv(t),\rhoUzp(t)]+\mD'[ \rhoUzp(t)]$,
where $\rhoUzp(t)\equiv\hUzp\hrho(t+T/2)\hUzpd$,
$\mD'$ is defined by $\lind_{ij}\to \lind_{ij}'=\hUzp \lind_{ij}\hUzpd$ in $\mD$,
and we have used Eq.~\eqref{eq:DS1}.
In fact, $\mD'=\mD$ holds true
because the time-independent part $\hnv^{0}$ of $\hnv(t)$ is invariant under $\hUzp$: $[\hUzp,\hnv^0]=0$
and hence the energy eigenstates $\ket{E_k}$ are the simultaneous eigenstates for $\hnv^{0}$ and $\hUzp$
(recall that $\lind_{ij}$ appears together with $\lind_{ij}^\dag$ in $\mD$).
Therefore, we have 
\begin{align}
	\partial_t\rhoUzp (t) =-i[\hnv(t),\rhoUzp (t)]+\mD[\rhoUzp (t)],
\end{align}
which is the same as Eq.~\eqref{eq:lindorg}.
As is the case in the high-frequency expansion,
we assume that Eq.~\eqref{eq:lindorg} leads to the unique time-periodic NESS $\rhoss(t)=\rhoss(t+T)$ in $t\gg\gamma^{-1}$.
%up to some time shift.
Then we have
\begin{align}
	\rhoss(t)
	=\rhossUzp (t) = \hUzp \rhoss(t+T/2)\hUzpd.
\end{align}
From this equation, we have the one-cycle average of an observable in Eq.~\eqref{eq:obsUodd} as
\begin{align}
	\bar{A} = \int_0^T \frac{dt}{T} \tr[\rhoss(t)\hO]
	=\int_0^T \frac{dt}{T} \tr[\rhoss(t+T/2)\hUzpd\hO\hUzp]
	=\int_0^T \frac{dt}{T} \tr[\rhoss(t)(-\hO)]
	=-\bar{A},
\end{align}
which means $\bar{A}=0$ for the NESS.
To obtain this, we have used,
the cyclic property of trace,
the periodicity of $\rhoss(t)$,
and Eq.~\eqref{eq:obsUodd}.

%###############################
Second,
we show that 
those one-cycle averages
vanish in our formula [Eq.~\eqref{eq:MR}] as well.
Recall that the micromotion part $\sigmm(t)$ does not contribute
and neither $\rhocan$ nor $\sigfe$ depends on time.
Thus we are to prove $\tr [\rhocan \hO]=\tr [\sigfe \hO ]=0$.
The first equation $\tr [\rhocan \hO]=0$ follows from 
the invariance of the static Hamiltonian $[\hUzp,\hnv^0]=0$ and Eq.~\eqref{eq:obsUodd}.
To show the second one $\tr [\sigfe \hO ]=0$,
we translate the dynamical symmetry [Eq.~\eqref{eq:DS1}] into the Fourier components:
\begin{align}
	(-1)^m \hUz_\pi\hH_m \hUzd_\pi=\hH_m,
\end{align}
which is obtained by Fourier-expanding both sides
of Eq.~\eqref{eq:DS1}.
This relation implies that
the effective Hamiltonian is invariant under
the unitary transformation: $\hUzp \Heff \hUzpd=\Heff$
and hence $\hUzp \dHeff \hUzpd=\dHeff$.
In fact, this relation leads to the invariance of the Floquet-engineering part $\sigfe$:
\begin{align}\label{eq:invfe}
\hUzp\sigfe\hUzpd=\sigfe	.
\end{align}
To show Eq.~\eqref{eq:invfe},
we compare the matrix elements
in the energy eigenbasis.
This basis is convenient because
$\hUzp\ket{E_k}=e^{i\theta_k}\ket{E_k}$
holds true.
The left-hand side of Eq.~\eqref{eq:invfe} gives
\begin{align}
\braket{E_k| \hUzp\sigfe\hUzpd|E_l}
&= e^{i\theta_k}\braket{E_k|\sigfe|E_l}e^{-i\theta_l}
= \frac{e^{i\theta_k}\braket{E_k|\dHeff|E_l}e^{-i\theta_l}}{(E_k-E_l)-i\gamma_{kl}}(\canw^{(k)}-\canw^{(l)}) 
\nonumber\\
&= \frac{\braket{E_k|\hUzp\dHeff\hUzpd|E_l}}{(E_k-E_l)-i\gamma_{kl}}(\canw^{(k)}-\canw^{(l)})
= \frac{\braket{E_k|\dHeff|E_l}}{(E_k-E_l)-i\gamma_{kl}}(\canw^{(k)}-\canw^{(l)})=\braket{E_k|\sigfe|E_l},
\end{align}
which thus equals the right-hand side of Eq.~\eqref{eq:invfe}.
Thus Eq.~\eqref{eq:invfe} has been proved
and leads to $\tr[\sigfe \hO]=0$
together with Eq.~\eqref{eq:obsUodd}.
Therefore, the one-cycle averages
for the observables in Eq.~\eqref{eq:obsUodd}
vanish in our formula~\eqref{eq:MR}.

%###############
Finally,
we show that
the one-cycle averages for those observables
vanish in the FGS.
In fact, a stronger statement holds true:
The one-cycle average vanishes for each Floquet state,
\begin{align}
	\int_0^T \frac{dt}{T} \braket{u_i(t)|\hO|u_i(t)}
	=\int_0^T \frac{dt}{T} \tr[\ket{u_i(t)}\bra{u_i(t)} \hO]
	=0.
\end{align}
Thanks to Eq.~\eqref{eq:obsUodd},
it is sufficient to show that
the one-cycle-averaged Floquet state
\begin{align}
	\rhofs_i \equiv \int_0^T\frac{dt}{T}\ket{u_i(t)}\bra{u_i(t)}.
\end{align}
is invariant under $\hUzp$ for each $i$.
This invariance follows from the dynamical symmetry~\eqref{eq:DS1} as follows.
Let us remember the defining equation of the Floquet state
\begin{align}
	\left[\hnv(t)-i\frac{d}{dt}\right]\ket{u_i(t)} = \epsilon_i \ket{u_i(t)}.
\end{align}
By applying $\hUzp$ from left,
shifting time as $t\to t+T/2$,
and making use of the dynamical symmetry~\eqref{eq:DS1},
we have
\begin{align}
	\left[\hnv(t)-i\frac{d}{dt}\right]\hUzp\ket{u_i(t+T/2)} = \epsilon_i \hUzp\ket{u_i(t+T/2)}.
\end{align}
Thus $\hUzp\ket{u_i(t+T/2)}$
is also the Floquet state
with quasienergy $\epsilon_i$.
Assuming that the quasienergies are not degenerate,
we obtain
\begin{align}
	\hUzp\ket{u_i(t+T/2)} = e^{i\varphi_i}\ket{u_i(t)}
\end{align}
for some phase $\varphi_i$.
Noticing the periodicity $\ket{u_i(t+T)}=\ket{u_i(t)}$,
we obtain
\begin{align}
	\rhofs_i = \int_0^T\frac{dt}{T}\ket{u_i(t)}\bra{u_i(t)}
	= \int_0^T\frac{dt}{T}\hUzp\ket{u_i(t+T/2)}\bra{u_i(t+T/2)}\hUzpd
	= \hUzp \rhofs_i \hUzpd,
\end{align}
which means $\rhofs_i$ is invariant under the unitary transform $\hUzp$
and hence $\tr[\rhofs_i \hO]=0$.
By taking the weighted average with $\fgw^{(i)}=e^{-\beta\epsilon_i}/Z_\text{FG}$,
we obtain
\begin{align}
	\int_0^T \frac{dt}{T}\tr[\fgs(t)\hO]
	=\sum_i \fgw^{(i)} \tr[\rhofs_i \hO] =0
\end{align}
for $\hO$ in Eq.~\eqref{eq:obsUodd}.
We note that, by replacing $\fgw^{(i)}$ by $\canw^{(i)}$, we obtain %a similar 
the same-type equation for the canonical Floquet steady state.

%#######
\subsection{Nonvanishing one-cycle averages}
We have shown that
the one-cycle averages for the four observables
in Eq.~\eqref{eq:obsUodd}
vanish for the actual dynamics,
our formulas [Eqs.~(8)--(10)], and the FGS, respectively.
In other words,
our formulas and the FGS
both respect the dynamical symmetry~\eqref{eq:DS1}
and give precise descriptions
for these observables.

\begin{figure*}[h]
	\includegraphics[width=\columnwidth]{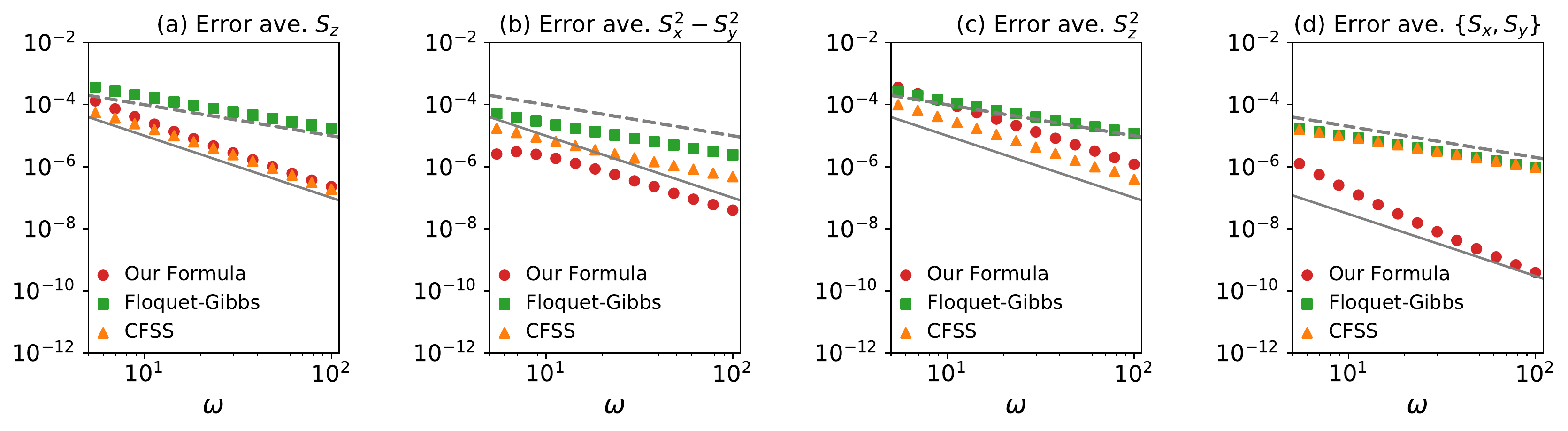}
	\caption{Difference of the one-cycle average
	calculated from the actual dynamics
	and that from our formula [Eq.~\eqref{eq:MR}] (circle), the FGS (square),
	and the CFSS (triangle)
	plotted against the driving frequency $\omega$.
	Each panel shows the result for the observables
	as described in the panel title.
%	(a) $\hSz$, (b) $\hSx^2-\hSy^2$, (c) $\hSz^2$, and (d) $\{\hSx,\hSy\}$.
	The solid and dashed lines are the guides to the eye	
	showing the lines with slopes $-2$ and $-1$, respectively.
	}
	\label{sfig:ave_vs}
\end{figure*}

Thus, for the complete comparison,
we are to discuss the remaining four observables:
$\hSz,\hSx^2-\hSy^2,\hSz^2$, and $\{\hSx,\hSy\}$.
In Fig.~\ref{sfig:ave_vs},
we plot the deviation
of the one-cycle average
calculated by our formula
and the FGS (as well as the canonical Floquet steady state for future reference)
from that of the actual dynamics.
While the deviation of the FGS is $\oo$
for all these observables,
that of our formula is $\ot$.
Thus our formula correctly describes
all the observables at $\oo$.

\begin{comment}
%####
In addition, for the special case of $E=0$,
the dynamical symmetry is extended as
$\hUzt\hnv(t+\frac{\theta{T}}{2\pi})\hUztd=\hnv(t)$ for any $\theta$.
This is why the oscillation was absent
in Fig.~\ref{fig:approach}(b)~\footnote{FN}.
More importantly, by using the symmetry for $\theta=\pi/4$,
we obtain that the one-cycle average vanishes
for $\hSx^2 - \hSy^2$ and $\hSx\hSy+\hSy\hSx$.

%#####
The above arguments mean that
the comparison of the CFSS and FGS
is completed by only considering 
$\hSz^2$, $\hSx^2 - \hSy^2$,
and $\hSx\hSy+\hSy\hSx$ for $E\neq0$,
and $\hSz^2$ for $E=0$.
\end{comment}

%#############################################
\subsection{One-cycle standard deviations}
\begin{figure*}[h]
	\includegraphics[width=\columnwidth]{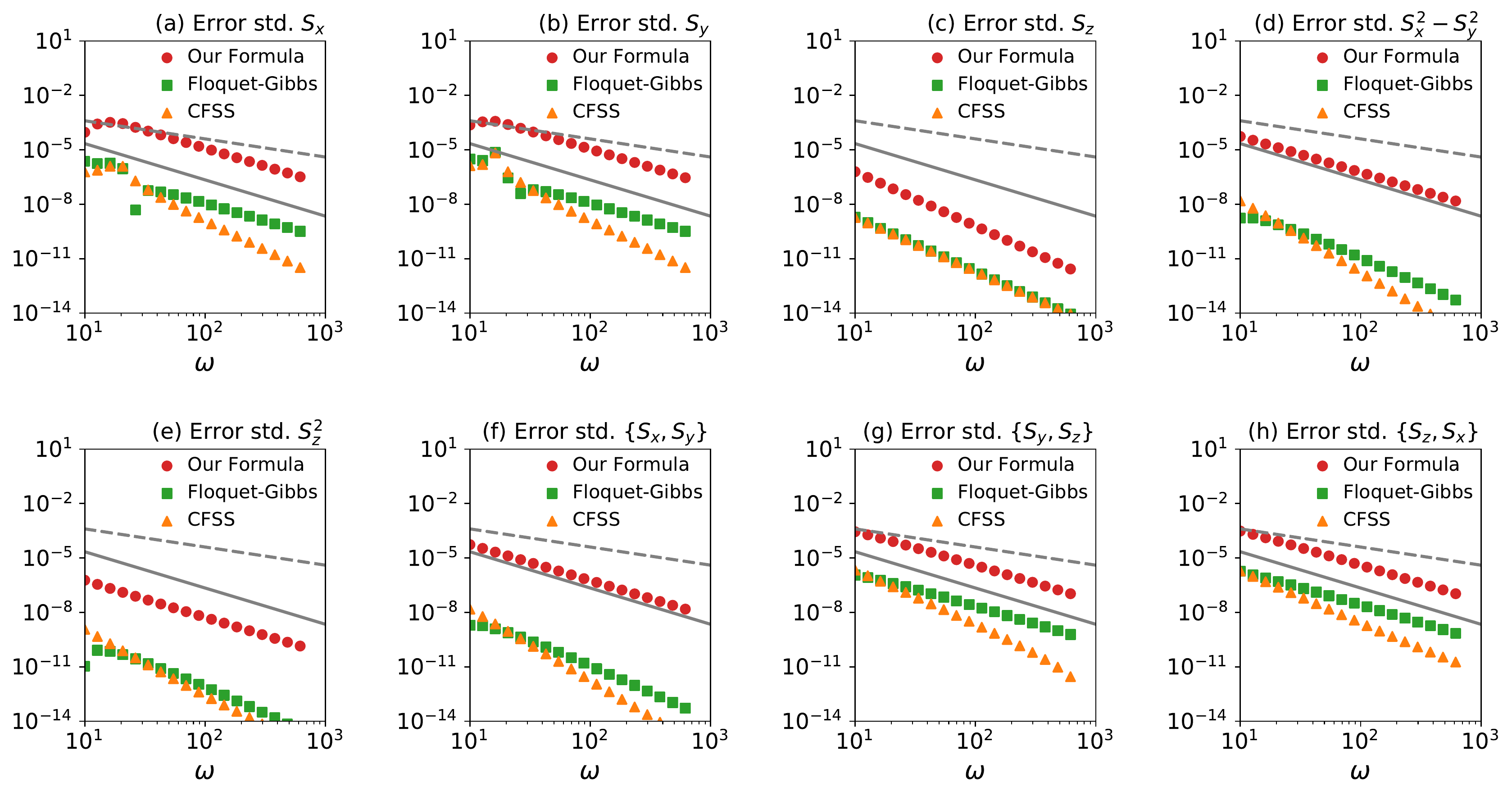}
	\caption{Difference of the one-cycle standard deviation
	calculated from the actual dynamics
	and that from our formula [Eq.~\eqref{eq:MR}] (circle), the FGS (square),
	and the CFSS (triangle)
	plotted against the driving frequency $\omega$.
	Each panel (a-h) shows the result for the observables
	as described in the panel title.
	The solid and dashed lines are the guides to the eye
	showing the lines with slopes $-2$ and $-1$, respectively.
	}
	\label{sfig:std_vs}
\end{figure*}

In the paper,
we have discussed the difference
of the one-cycle standard deviation $\Delta\Sigma_A(\omega)$
for the representative two observables
$\hO=\hSz$ and $\{\hSx,\hSy\}$.
Here we supplement the data,
plotting $\Delta\Sigma_A(\omega)$
for all the eight observables
calculated with our formula [Eqs.~(8) and (9)],
the FGS (as well as the CFSS
for future reference)
in Fig.~\ref{sfig:std_vs}.

%######
The difference $\Delta\Sigma_A(\omega)$ between the actual dynamics and our formula
is $\ot$ for all observables as shown in Fig.~\ref{sfig:std_vs}.
This result supports that our micromotion part $\sigmm(t)$
properly describes the NESS at $\oo$.
Quantitatively, $\Delta\Sigma_A(\omega)$ is smaller for the FGS,
where all-order contributions in $\omega^{-1}$ are included.
We could improve the accuracy of our formula by extending our formula to higher orders.

%##############################
\section{Breakdown of antiunitary dynamical symmetry}
We supplement the argument in the paper
that the one-cycle average of $\hO=\{\hSx,\hSy\}$
vanishes for the FGS
but does not for the actual dynamics and our formulas [Eqs.~(8)--(10)].
In the paper,
we have shown that the antiunitary operator $\hat{V}$
and the associated dynamical symmetry
\begin{align}\label{eq:DS2}
\hat{V}\hnv(T-t)\hat{V}^\dag=\hnv(t)	
\end{align}
lead to the vanishing one-cycle average
for the FGS.
Let us see how such an antiunitary
dynamical symmetry does not constrain
the actual dynamics or our formula
due to dissipation.

%####
First, we discuss the actual dynamics described
by the Lindblad equation~\eqref{eq:lindorg}.
To utilize the antiunitary dynamical symmetry,
we substitute $t$ by $T-t$ 
and multiply $\hat{V}$ from left and $\hat{V}^\dag$ from right.
Then, we have
\begin{align}\label{eq:lindS1}
		-\partial_t\rhoV (t) =i[\hnv(t),\rhoV (t)]+\mD''[\rhoV (t)],
\end{align}
where we have used Eq.~\eqref{eq:DS2},
$\rhoV(t) \equiv \hat{V}\hrho(T-t)\hat{V}^\dag$,
and
$\mD''$ is defined by $\lind_{ij}\to \lind_{ij}''=\hat{V} \lind_{ij}\hat{V}^\dag$ in $\mD$.
We notice that $\mD''=\mD$
because the time-independent Hamiltonian $\hnv^0$ is invariant under the antiunitary transform $\hat{V}$
similarly to the argument in Sec.~\ref{sec:vanishingave}.
Therefore, Eq.~\eqref{eq:lindS1} leads to
\begin{align}
	\partial_t\rhoV (t) =-i[\hnv(t),\rhoV (t)]-\mD[\rhoV (t)].
\end{align}
We note that the sign of the $\mD$ term
has changed from the original Lindblad equation~\eqref{eq:lindorg}
and $\rhoV(t)$ cannot be related directly to $\hrho(t)$.
Thus the antiunitary dynamical symmetry~\eqref{eq:DS2}
does not constrain the actual dynamics
in the presence of dissipation.

%####
Second,
we show that our formula is not constrained by
the antiunitary dynamical symmetry~\eqref{eq:DS2}.
More concretely,
we have $\hat{V}\sigfe\hat{V}^\dag\neq\sigfe$
unlike the case of unitary transformations.
To show this, we first notice that
the dynamical symmetry~\eqref{eq:DS2}
leads to $\hat{V}\hH_m\hat{V}^\dag=\hH_m$
for the Fourier components
and to $\hat{V}\Heff\hat{V}^\dag=\Heff$ and hence $\hat{V}\dHeff\hat{V}^\dag=\dHeff$.
We second notice $\hat{V}=K\hUzp$,
where $K$ is the complex conjugate operator.
Then, we consider the matrix elements
of $\hat{V}\sigfe\hat{V}^\dag$
in the energy eigenbasis:
\begin{align}
	\braket{E_k|\hat{V}\sigfe\hat{V}^\dag|E_l}
	&=e^{i\theta_k}\braket{E_k|K\sigfe K|E_l}e^{-i\theta_l}
	=e^{i\theta_k}\braket{E_k|\sigfe |E_l}^*e^{-i\theta_l}
\nonumber\\
	&=\frac{e^{i\theta_k}\braket{E_k|\dHeff|E_l}^*e^{-i\theta_l}}{(E_k-E_l)+i\gamma_{kl}}(\canw^{(k)}-\canw^{(l)})
	=\frac{\braket{E_k|\hUzpd\dHeff\hUzp|E_l}^*}{(E_k-E_l)+i\gamma_{kl}}(\canw^{(k)}-\canw^{(l)})
\nonumber\\
	&=\frac{\braket{E_k|\dHeff|E_l}^*}{(E_k-E_l)+i\gamma_{kl}}(\canw^{(k)}-\canw^{(l)})
	\neq \braket{E_k|\sigfe|E_l}.
\end{align}
Although $\braket{E_k|\dHeff|E_l}^*=\braket{E_k|\dHeff|E_l}$ in fact,
the sign of $\gamma_{kl}$ has changed from $\braket{E_k|\dHeff|E_l}$.
Thus, in the presence of dissipation,
$\hat{V}\sigfe\hat{V}^\dag\neq\sigfe$
and $\tr(\sigfe\hO)\neq0$ in general
even if $\hat{V}\hO\hat{V}^\dag=-\hO$.

%##############################
\section{Canonical Floquet Steady State (CFSS)}
Here we introduce the canonical Floquet steady state (CFSS)
\begin{align}\label{seq:cfss}
	\cfss (t) = \frac{1}{Z}\sum_i e^{-\beta E_i} \ket{u_i(t)}\bra{u_i(t)}
	= \frac{1}{Z}\sum_i e^{-\beta E_i} \ket{\psi_i(t)}\bra{\psi_i(t)},
\end{align}
where $Z=\sum_i e^{-\beta E_i}$, $\ket{u_i(t)}$ is the Floquet state,
and $\ket{\psi_i(t)}=e^{-i\epsilon_i t}\ket{u_i(t)}$
is the corresponding solution of the time-dependent Schr\"{o}dinger equation
with $\epsilon_i$ being the quasienergy.
Here, we have assumed that the driving frequency $\omega$ is so large
and $\ket{u_i(t)}$ is so close to $\ket{E_i}$
that
%the labels of $E_i$ and $\ket{u_i(t)}$ are well-defined by $\ket{u_i(t)}\simeq \ket{E_i}$
there is the one-to-one correspondence between $\ket{E_i}$ and $\ket{u_i(t)}$ for each index $i$.

%#####
The difference between the FGS and CFSS is the weight factor.
This is defined by the quasienergy $\epsilon_i$ for the FGS
whereas by the real energy $E_i$ for the CFSS.
This difference is quantitatively important
because $E_i-\epsilon_i =\oo$
and the FGS and CFSS can give different scalings in $\omega$
at high frequency.

%#####
The difference of the one-cycle-averaged observables
calculated for the actual dynamics and the CFSS
is shown in Fig.~\ref{sfig:ave_vs}.
For the two observables $\hSz$ and $\hSz^2$,
the CFSS gives the appropriate $\omega^{-2}$ scaling
which is not captured by the FGS.
For the other two $\hSx^2-\hSy^2$ and $\{\hSx,\hSy\}$,
the CFSS deviates from the actual value at $\oo$
and fails to describe the actual dynamics at $\oo$.
The CFSS thus provide partly improved descriptions for some observables than the FGS.
It is noteworthy that
the CFSS does not involve any information about the system-bath coupling like the FGS.

%######
The difference of the one-cycle standard deviations $\Delta\Sigma_A(\omega)$
calculated for the actual dynamics and the CFSS
is shown in Fig.~\ref{sfig:std_vs}.
At high-frequency, the CFSS leads to more rapid decreases of $\Delta\Sigma_A(\omega)$
than the FGS for most observables.
Thus the CFSS gives improved descriptions of the NESS than the FGS.

%##############################
\section{Equivalence of our formula and CFSS in $\Gamma_{ij}\to0$}
Here we show that, in the weak dissipation limit $\Gamma_{ij}\to0$,
our formula [Eqs.~(8)--(10)] coincides with the CFSS
rather than the FGS.
Since the extension to the degenerate case is straightforward,
we consider the case where $\hzero$ is nondegenerate for simplicity.

The weak dissipation limit of our formula is obtained
just by replacing $\gamma_{ij}$ with 0 in $\sigfe$:
\begin{align}
\hrho(t)= \rhocan + \sigmm(t) + \sigfe + O(\omega^{-2}),\label{eq:s7:MR}
\end{align}
with
\begin{align}
	\braket{E_k|\sigfe|E_l} =\frac{\braket{E_k|\dHeff|E_l}}{E_k-E_l}(\canw^{(k)}-\canw^{(l)}) \qquad (k\neq l)
\end{align}
and $\braket{E_k|\sigfe|E_k}=0$.
We will show that $\cfss(t)$ coincides with the above $\hrho(t)$
by considering its high-frequency expansion.

%###
This is achieved by finding the solution $\ket{\psi_i(t)}$
within the high-frequency expansion.
According to Ref.~\cite{EckardtA15},
$\ket{\psi_k(t)}$ can be represented as
\begin{align}
	\ket{\psi_k(t)} = e^{G(t)}\ket{\psi_k(0)},
\end{align}
where
\begin{align}
	G(t) = \frac{1}{\omega} \sum_{m\neq0}\frac{e^{-im\omega t}}{m}\hH_m +\ot,
\end{align}
and $\ket{\psi_k(0)}$ is the eigenstate of the effective Hamiltonian $\Heff=\hzero+\dHeff$ with eigenvalue $\epsilon_k = E_k +\oo$.
Since $\dHeff=\oo$ as shown in Sec.~\ref{app:HFexpansion},
$\ket{\psi_k(0)}$ can be obtained by the standard perturbation technique as
\begin{align}
	\ket{\psi_k(0)} = \ket{E_k} + \sum_{l(\neq k)} \ket{E_l}\frac{\braket{E_l|\dHeff|E_k}}{E_k-E_l}+\ot.
\end{align}
\begin{comment}
This equation leads to
\begin{align}
	\ket{\psi_k(0)}\bra{\psi_k(0)}
	&=\ket{E_k}\bra{E_k} +  \sum_{l(\neq k)}\ket{E_k} \frac{\braket{E_k|\dHeff|E_l}}{E_k-E_l}\bra{E_l} 
	+  \sum_{l(\neq k)} \ket{E_l}\frac{\braket{E_l|\dHeff|E_k}}{E_k-E_l}\bra{E_k} +\ot\\
	&=\ket{E_k}\bra{E_k} +  \sum_{l(\neq k)}\ket{E_k} \frac{\braket{E_k|\dHeff|E_l}}{E_k-E_l}\bra{E_l} 
	+  \sum_{l(\neq k)} \ket{E_l}\frac{\braket{E_l|\dHeff|E_k}}{E_k-E_l}\bra{E_k} +\ot
\end{align}
\end{comment}
Substituting this equation into Eq.~\eqref{seq:cfss},
we obtain
\begin{align}
	&\cfss (t) \notag\\
	&=\sum_k \canw^{(k)} e^{G(t)} \left[ \ket{E_k}\bra{E_k}
	+  \sum_{l(\neq k)} \frac{\braket{E_k|\dHeff|E_l}}{E_k-E_l}\ket{E_k}\bra{E_l} 
	+  \sum_{l(\neq k)} \frac{\braket{E_l|\dHeff|E_k}}{E_k-E_l}\ket{E_l}\bra{E_k}   \right] e^{-G(t)}+\ot
\nonumber\\
	&=e^{G(t)}\rhocan e^{-G(t)} +\sum_{\substack{k,l\\(k\neq l)}} \left[ \canw^{(k)} \frac{\braket{E_k|\dHeff|E_l}}{E_k-E_l}\ket{E_k}\bra{E_l}
	+
	\canw^{(k)}\frac{\braket{E_l|\dHeff|E_k}}{E_k-E_l}\ket{E_l}\bra{E_k} 
	\right]+\ot
\nonumber\\
	%&=\rhocan +\sigmm(t) +\sum_{\substack{k,l\\(k\neq l)}} (\canw^{(k)}-\canw^{(l)}) \frac{\braket{E_k|\dHeff|E_l}}{E_k-E_l}\ket{E_k}\bra{E_l}+\ot \nonumber\\
	&=e^{\mG(t)}\rhocan  +\sum_{\substack{k,l\\(k\neq l)}} (\canw^{(k)}-\canw^{(l)}) \frac{\braket{E_k|\dHeff|E_l}}{E_k-E_l}\ket{E_k}\bra{E_l}+\ot \nonumber\\
	&= \rhocan+\sigmm(t) +\sigfe +\ot,
\end{align}
which is equal to our formula~\eqref{eq:s7:MR}
($\mG(t)$ was defined in Sec.~\ref{app:HFexpansion}).
We note that the FGS deviates from the CFSS in general by $\oo$ since $E_i-\epsilon_i=\oo$.
Thus, in the small dissipation limit, the NESS coincides with the CFSS rather than the FGS.

\end{document}